\documentclass[10pt,twocolumn,letterpaper]{article}

\usepackage[pagenumbers]{cvpr} 
\usepackage{url}
\usepackage{adjustbox}
\usepackage{tikz}
\usepackage{amssymb}
\usepackage{wrapfig}
\usepackage{enumitem}
\usepackage{marvosym}
\usepackage{multirow}
\usepackage{amsmath}
\usepackage{subcaption}

\usepackage[dvipsnames,table]{xcolor}
\usepackage[most]{tcolorbox}

\tcbset{
  mycontribbox/.style={
    colback=blue!5!white,
    colframe=blue!45!black!65,
    boxrule=0.8pt,
    arc=3pt,
    left=6pt,
    right=6pt,
    top=4pt,
    bottom=4pt,
    fonttitle=\bfseries,
  }
}

\definecolor{cvprblue}{rgb}{0.21,0.49,0.74}
\usepackage[pagebackref,breaklinks,colorlinks,allcolors=cvprblue]{hyperref}
\usepackage[capitalize]{cleveref}

\newcommand{\cross}{\textcolor{teal}{\Crossedbox}}
\renewcommand{\check}{\textcolor{red}{\Checkedbox}}

\newcommand\blfootnote[1]{%
  \begingroup
  \renewcommand\thefootnote{}\footnote{#1}%
  \addtocounter{footnote}{-1}%
  \endgroup
}

\def\paperID{11469} 
\def\confName{CVPR}
\def\confYear{2026}

\title{MedNeXt-v2: Scaling 3D ConvNeXts for Large-Scale Supervised Representation Learning in Medical Image Segmentation}

\author{Saikat Roy$^{*,1, 2}$, Yannick Kirchhoff$^{*,1,2,3}$, Constantin Ulrich$^{1,4,5}$, Maximillian Rokuss$^{1,2}$,\\Tassilo Wald$^{1,2,6}$, Fabian Isensee$^{1,6}$, Klaus Maier-Hein$^{1,2,7}$\\\\
$^{1}$German Cancer Research Center (DKFZ) Heidelberg, Division of Medical Image\\
Computing, Germany\\
$^{2}$Faculty of Mathematics and Computer Science, Heidelberg University, Germany\\
$^{3}$HIDSS4Health - Helmholtz Information and Data Science School for Health,\\
Karlsruhe/Heidelberg, Germany\\
$^{4}$Medical Faculty Heidelberg, Heidelberg University, Germany\\
$^{5}$National Center for Tumor Diseases (NCT), Heidelberg, Germany\\
$^{6}$Helmholtz Imaging, German Cancer Research Center, Germany\\
$^{7}$Pattern Analysis and Learning Group, Department of Radiation Oncology, \\Heidelberg University Hospital, Germany\\
{\tt\small saikat.roy@dkfz-heidelberg.de; yannick.kirchhoff@dkfz-heidelberg.de}
}

\begin{document}

\maketitle

\begin{abstract}
Large-scale supervised pretraining is rapidly reshaping 3D medical image segmentation. However, existing efforts focus primarily on increasing dataset size and overlook the question of whether the backbone network is an effective representation learner at scale. In this work, we address this gap by revisiting ConvNeXt-based architectures for volumetric segmentation and introducing MedNeXt-v2, a compound-scaled 3D ConvNeXt that leverages improved micro-architecture and data scaling to deliver state-of-the-art performance. First, we show that routinely used backbones in large-scale pretraining pipelines are often suboptimal. Subsequently, we use comprehensive backbone benchmarking prior to scaling and demonstrate that stronger from scratch performance reliably predicts stronger downstream performance after pretraining. Guided by these findings, we incorporate a 3D Global Response Normalization module and use depth, width, and context scaling to improve our architecture for effective representation learning. We pretrain MedNeXt-v2 on 18k CT volumes and demonstrate state-of-the-art performance when fine-tuning across six challenging CT and MR benchmarks (144 structures), showing consistent gains over seven publicly released pretrained models. Beyond improvements, our benchmarking of these models also reveals that stronger backbones yield better results on similar data, representation scaling disproportionately benefits pathological segmentation, and that modality-specific pretraining offers negligible benefit once full finetuning is applied. In conclusion, our results establish MedNeXt-v2 as a strong backbone for large-scale supervised representation learning in 3D Medical Image Segmentation. Our code and pretrained models are made available with the official nnUNet repository at: \url{https://www.github.com/MIC-DKFZ/nnUNet}.
\blfootnote{*Contributed equally. Each author may denote themselves as positional first author in their CVs.}
\end{abstract}


\section{Introduction}
\label{sec:introduction}
Automated segmentation of medical images is one of the most common tasks in biomedical image analysis \cite{maier2018rankings, nestle2005comparison,de2018clinically,hollon2020near}. Despite rapid development in deep learning based approaches over the last decade \cite{minaee2021image,abdou2022literature,litjens2017survey}, UNet-based \cite{ronneberger2015u} deep convolutional networks (ConvNets) have remained central to high-performing methodologies for 3D medical image segmentation \cite{isensee2021nnu,isensee2024nnu}. Although alternative approaches such as Transformers have been popular in recent years \cite{khan2025recent}, their limited inductive bias has proved a hindrance for training from scratch on the currently available medical segmentation datasets, typically containing sparse annotations \cite{litjens2017survey}. This has led to ConvNeXt-based \cite{liu2022convnet} approaches leveraging the scalability of the Transformer while retaining the strong inductive bias of ConvNets to offer effective solutions for 3D medical image segmentation \cite{roy2023mednext,liu2024fast,lee20223d,rahman2025efficientmednext,cao2024hdnext}.

\begin{figure}
    \centering
    \includegraphics[width=\linewidth, trim=10 20 10 10, clip]{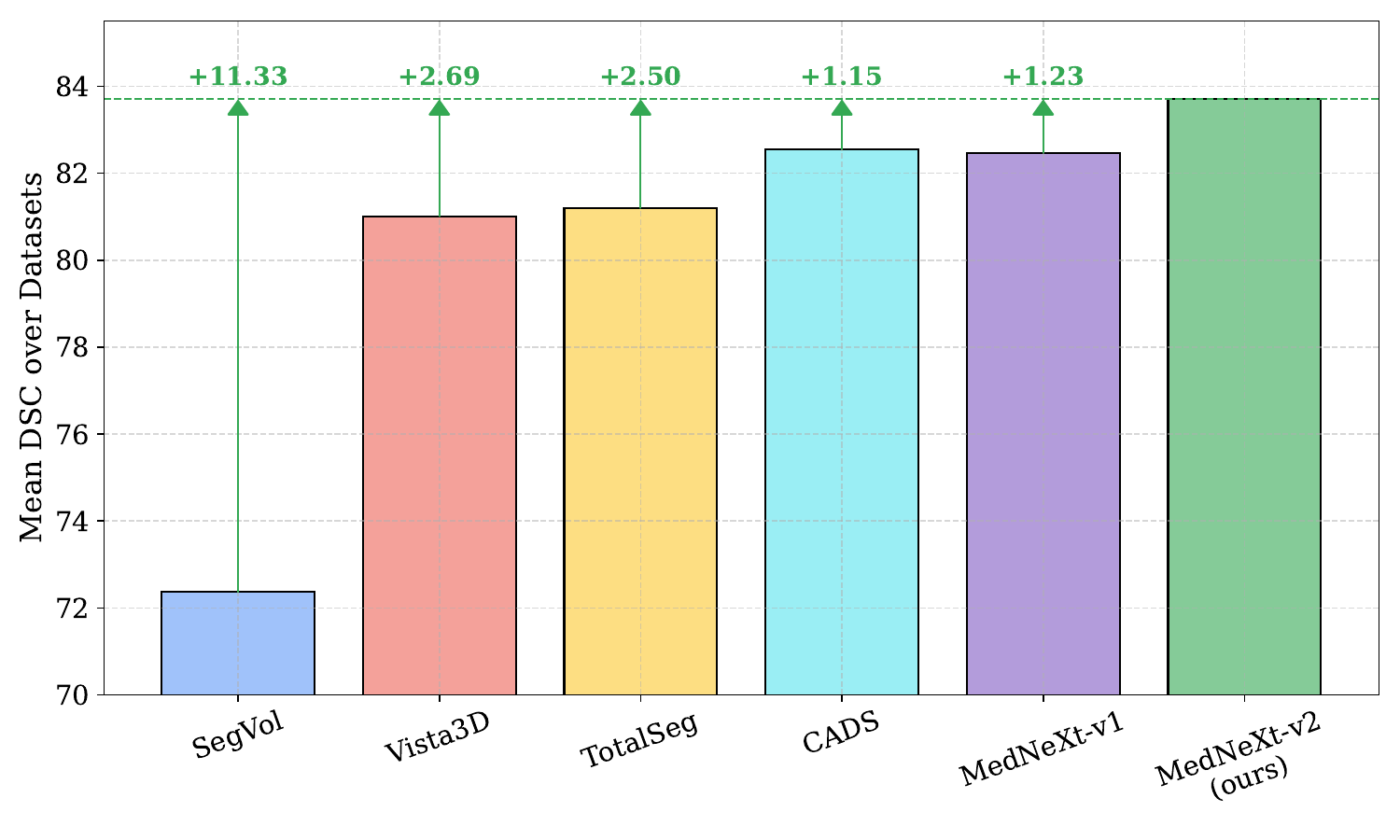}
    \caption{\textbf{MedNeXt-v2 sets a new state-of-the-art in 3D medical image segmentation.} By leveraging micro-architectural improvements and large-scale pretraining, it outperforms powerful existing networks across multiple 3D medical segmentation tasks.}
    \label{fig:fig1}
\end{figure}

However, following significant advances in computer vision \citep{sun2017revisiting,radford2021learning,dehghani2023scaling} over the last decade, the field of medical image segmentation has also been gradually moving towards incorporating large-scale supervised pretraining of deep networks \citep{butoi2023universeg,ulrich2023multitalent,wang2024sam,he2025vista3d}. In recent years, the availability of large monolithic datasets \citep{wasserthal2023totalsegmentator,akinci2025totalsegmentator} or aggregated collections of previously available small-scale datasets \citep{li2024abdomenatlas,bai2024m3d,xu2025cads} has led to initial attempts at pretraining large-scale deep learning models for the segmentation of 3D medical images. Notably, while approaches in 2D computer vision have moved towards self-supervised learning (SSL) owing to the abundance of unlabeled data \cite{jing2020self,gui2024survey}, the domain of 3D medical image segmentation continues to leverage supervised pretraining. Despite some initial SSL studies \cite{wald2025openmind,wald2025revisiting} with 3D medical image pretraining, the efficiency in learning representations in supervised pretraining continues be of interest. For example, in \cite{li2025well}, it was demonstrated that supervised representation learning is 99.6\% more data efficient than an SSL approach. However, annotated sample sizes remain limited compared to those of natural image datasets ($O(10^4)$ vs $O(10^7)$) \cite{roylost}, as the annotation of 3D medical images requires experienced radiologists and substantial costs arising from the substantial manual effort. Therefore, we consider \textit{only supervised pretraining} in the course of this work. While initial attempts have proven effective \citep{ulrich2023multitalent,li2025well,li2024abdomenatlas,rokuss2025lesionlocator} for learning generalized representations, our work leads us to systemic issues with supervised pretraining in 3D medical image segmentation.

\noindent\textbf{Challenges in supervised pretraining:} Initial efforts towards the supervised pretraining of deep networks for medical image analysis have a number of recurring issues as shown in \cref{tab:systemic} -- \textbf{P1)} \textbf{Legacy Backbones:} While large-scale datasets in natural image analysis are more established, such as ImageNet, MS-COCO, ADE20k \citep{deng2009imagenet,lin2014microsoft,zhou2017scene}, the construction of large medical image datasets is a vigorously active area of research \citep{qu2023abdomenatlas,xu2025cads,bourigault2025ukbob,graf2024totalvibesegmentator,akinci2025totalsegmentator}. Therefore, owing to dataset creation itself being an active line of research, many recent efforts share a limitation in solely emphasizing on dataset scale, often aiming to demonstrate the efficacy of their large pretraining corpora on \textit{any} deep architecture in general, but not necessarily combining it with a state-of-the-art network for 3D medical image segmentation (4 out of 11 cases). \textbf{P2)} \textbf{Scaling data \textit{after} backbone validation:} The preceding problem could, in theory, be eliminated by benchmarking the backbone architecture prior to pretraining. However, as seen in \cref{tab:systemic}, this is not usually performed (2 out of 11 cases) -- but as shown later in \cref{tab:main_results}, stronger backbones lead to stronger pretrained networks. \textbf{P3)} \textbf{Comparisons against only from-scratch baselines:} Finally, we observe that large-scale pretrained models are usually evaluated solely against models trained from scratch. While this does indeed highlight the benefits of pretraining in general, it does little to highlight the effectiveness of a \textit{specific} pretraining strategy. We observe that 4 out of 11 surveyed methods evaluate against other pretrained methods, leading to questions of effectivness of such methods as pretrained backbones.

In this work, we attempt to bridge this gap by shifting the focus from dataset size alone to representation quality by asking: \textcolor{violet}{``\textit{Is our large-scale representation learner suitable for learning effective representation from dataset scaling?}"}. In doing so, we offer a ConvNeXt-based architecture for state-of-the-art 3D medical image segmentation. Our contributions can be expressed as combination of validating our network for effective small-scale segmentation performance \textit{prior to} pairing it with effective large-scale pretraining and downstream evaluation, as follows:

\begin{enumerate}[leftmargin=*]
    \item \textbf{Backbone Validation:} The standard approach of selecting networks for effective large-scale representation learning for medical images is seemingly selecting the best off-the-shelf network available based on external baselines \citep{wasserthal2023totalsegmentator,graf2024totalvibesegmentator,akinci2025totalsegmentator,xu2025cads} or using their own proposed network with limited small-scale evaluations \citep{huang2023stu}. A lack of focus in this direction \citep{bourigault2025ukbob} can also lead to the selection of methods for large-scale training, which have notably worse performance against state-of-the-art methods in multiple benchmarks \citep{bassi2024touchstone,isensee2024nnu}. To address this, we propose to actively benchmark our network on effective segmentation performance for multiple pathological and anatomical structures on \textit{multiple} small-scale 3D datasets. Subsequently, we demonstrate that benchmarking performance correlates strongly to downstream performance following pretraining. (\cref{sec:scaling-after})

    \item \textbf{Leveraging large-scale pretraining of ConvNeXts:} We introduce MedNeXt-v2, which is a ConvNeXt-based \cite{woo2023convnext} architecture, for state-of-the-art 3D medical image segmentation. However, we do not claim complicated architectural innovations. Instead, we leverage sensible micro-architectural improvements via a 3D Global Response Normalization (GRN) module and combine it with large-scale pretraining on 18k 3D CT volumes on 40 structures. To increase the efficacy of fine-tuning, we demonstrate that a simple increase in input context during a short fine-tuning phase is able to outperform 7 equivalently trained large-scale competitive baselines (some trained on more data, or larger size) on 6 challenging CT and MR downstream segmentation tasks on 144 structures as summarized in \cref{fig:fig1}. (\cref{sec:mednext-v2})

    \item \textbf{Benchmarking of large-scale supervised pretraining in medical image segmentation:} Finally, large-scale pretraining of 3D medical image segmentation networks is an relatively newer and active area of research. Accordingly, there is limited availability of benchmarks specifically for downstream fine-tuning to draw generalized conclusions for 3D supervised pretraining. In this work, we are the first to benchmark seven publicly available large-scale pretrained networks for 3D medical image segmentation and compare it against 3 networks proposed by us. We derive insights about the influence of pretraining backbones on downstream finetuning performance, the effectiveness of modality-specific pretraining and anatomy-specific performance gains. (\cref{sec:benchmarking_analysis})

\end{enumerate}

\begin{table}[ht]
    \centering
    \begin{adjustbox}{width=0.47\textwidth}
    \begin{tabular}{lrrccc}
    \toprule
    \textbf{Methods} & \textbf{Backbones} & \textbf{Venues} & \textbf{P1} & \textbf{P2} & \textbf{P3} \\
    \toprule \toprule
    \rowcolor{gray!20} \textbf{M01} \citep{wang2024sam} & ViT & ECCV'23 & \check & \cross & \check \\
    \textbf{M02} \citep{he2025vista3d} & SegResNet & CVPR'25 & \check & \cross & \cross \\
    \rowcolor{gray!20} \textbf{M03} \citep{wasserthal2023totalsegmentator} & nnUNet & Rad:AI'23 & \cross & \cross & \check \\
    \textbf{M04} \citep{akinci2025totalsegmentator} & nnUNet & Rad'25 & \cross & \cross & \check \\
    \rowcolor{gray!20} \textbf{M05} \citep{graf2024totalvibesegmentator} & ResEncL & EuroRad'25 & \cross & \cross & \check \\
    \textbf{M06} \citep{xu2025cads} & ResEncL & arXiv'25 & \cross & \cross & \check \\
    \rowcolor{gray!20} \textbf{M07} \citep{bourigault2025ukbob} & SwinUNETR & ICCV'25 & \check & \check & \check \\
    \textbf{M08} \citep{li2025well} & SwinUNETR & ICLR'24 & \check & \cross & \check \\
    \rowcolor{gray!20} \textbf{M09} \cite{silva2023towards} & SwinUNETR & MICCAI'23 & \check & \cross & \check \\
    \textbf{M10} \citep{liu2024universal} & SwinUNETR & MedIA'24 & \check & \check & \cross \\
    \textbf{M11} \citep{gao2024training} & (Res) UNet & ICCV'24 & \check & \cross & \cross \\
    \bottomrule
    \end{tabular}
    \end{adjustbox}

    \caption{\textbf{Systemic gaps in large-scale 3D medical image segmentation.} \textbf{P1: Legacy architectures.} Many large-scale studies default to popular baselines (e.g., nnU-Net), treating architecture as fixed and focusing the problem to ``more data,'' while underusing stronger representation learners. \textbf{P2: Scaling data \textit{after} Backbone validation.} Almost \textit{universally}, architectures are ported \textit{directly} to large-scale training with \textit{limited or no} small-scale benchmarking -- implicitly assuming that more data will yield strong performance. This assumption has no guarantees and excludes architecture that could have benefited more from pretraining. \textbf{P3: Only non-pretrained baselines.} Finally, large-scale pretrained models are sometimes evaluated only against baselines trained from scratch, thereby undercutting comparisons to \textit{true} state-of-the-art. Here, \check ~ represents the presence of a problem.}
    \label{tab:systemic}
\end{table}

\begin{table}[ht]
    \centering
    \begin{adjustbox}{width=0.475\textwidth}
    \begin{tabular}{rccccc}
    \toprule
    \textbf{Backbones} & \textbf{BTCV} & \textbf{AMOS} & \textbf{KITS} & \textbf{ACDC} & \textbf{AVG} \\
    \toprule \toprule
    \rowcolor{red!10} \multicolumn{6}{c}{\textbf{Existing backbones as in \cref{tab:systemic}}}\\
    \midrule
    \rowcolor{gray!20} \textbf{UNETR (ViT)} & 73.11 & 79.71 & 80.81 & 91.38 & 81.25 \\
    \textbf{SegResNet} & 80.22 & 86.83 & 83.34 & 92.38 & 85.69 \\
    \rowcolor{gray!20} \textbf{nnUNet} & 79.84 & 87.95 & 87.67 & 93.34 & 87.20 \\
    \textbf{SwinUNETR} & 77.20 & 85.39 & 85.09 & 92.59 & 85.07 \\
    \rowcolor{gray!20} \textbf{ResEnc-L} & 80.66 & \textbf{88.12} & 88.25 & \underline{93.44} & 87.62 \\
    \midrule
    \rowcolor{green!10} \multicolumn{6}{c}{\textbf{State of the Art backbones as in} \cite{isensee2024nnu,bassi2024touchstone}}\\
    \midrule
    \rowcolor{gray!20} \textbf{CoTr} & 78.17 & 87.05 & 85.39 & \textbf{93.47} & 86.02 \\
    \textbf{STU-Net} & 80.24 & 87.67 & 87.02 & 93.15 & 87.02 \\
    \rowcolor{gray!20} \textbf{MedNeXt-v1} & \textbf{81.02} & 88.02 & \underline{88.84} & 93.08 & \underline{87.74} \\
    \midrule
    \textbf{MedNeXt-v2} & \underline{80.99} & \underline{88.05} & \textbf{89.09} & 93.39 & \textbf{88.03} \\
    \bottomrule
    \end{tabular}
    \end{adjustbox}

    \caption{\textbf{Benchmarking \textit{before} Data Scaling.} While there is a limited history of benchmarking backbones \cite{gao2024training,bourigault2025ukbob} before dataset scaling, this step is usually ignored and as demonstrated, to potentially negative consequences. In fact, we demonstrate that 4 out of 5 architectures used in \cref{tab:systemic}, are outperformed by easy to train but better representation learners such as ResEncL or MedNeXt-v1 on 4 public datasets, with SOTA architectures derived from recent large-scale benchmarking studies for 3D medical image segmentation \cite{isensee2024nnu,bassi2024touchstone}. This ineffectiveness of using sub-par backbones for learning generalized representations, therefore, weakens the models paired with such large-scale 3D medical image datasets. \textbf{Bold}: Best results, \underline{Underline}: Second-best results.
    }
    \label{tab:benchmarking_backbones}
\end{table}





\section{Method}

\begin{figure}[h]
    \begin{subfigure}[b]{0.235\textwidth}
        \centering
        \includegraphics[width=\linewidth, trim=0.8cm 0.8cm 0.6cm 0.8cm, clip]{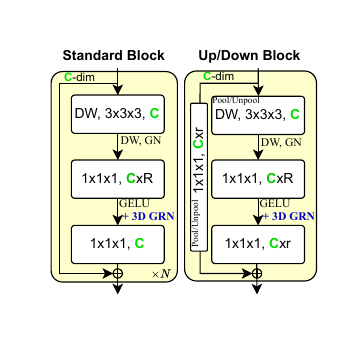}
        \caption{Micro-arch. of MedNeXt-v2}
        \label{fig:sub1}
    \end{subfigure}
    \begin{subfigure}[b]{0.235\textwidth}
        \centering
        \includegraphics[width=\linewidth]{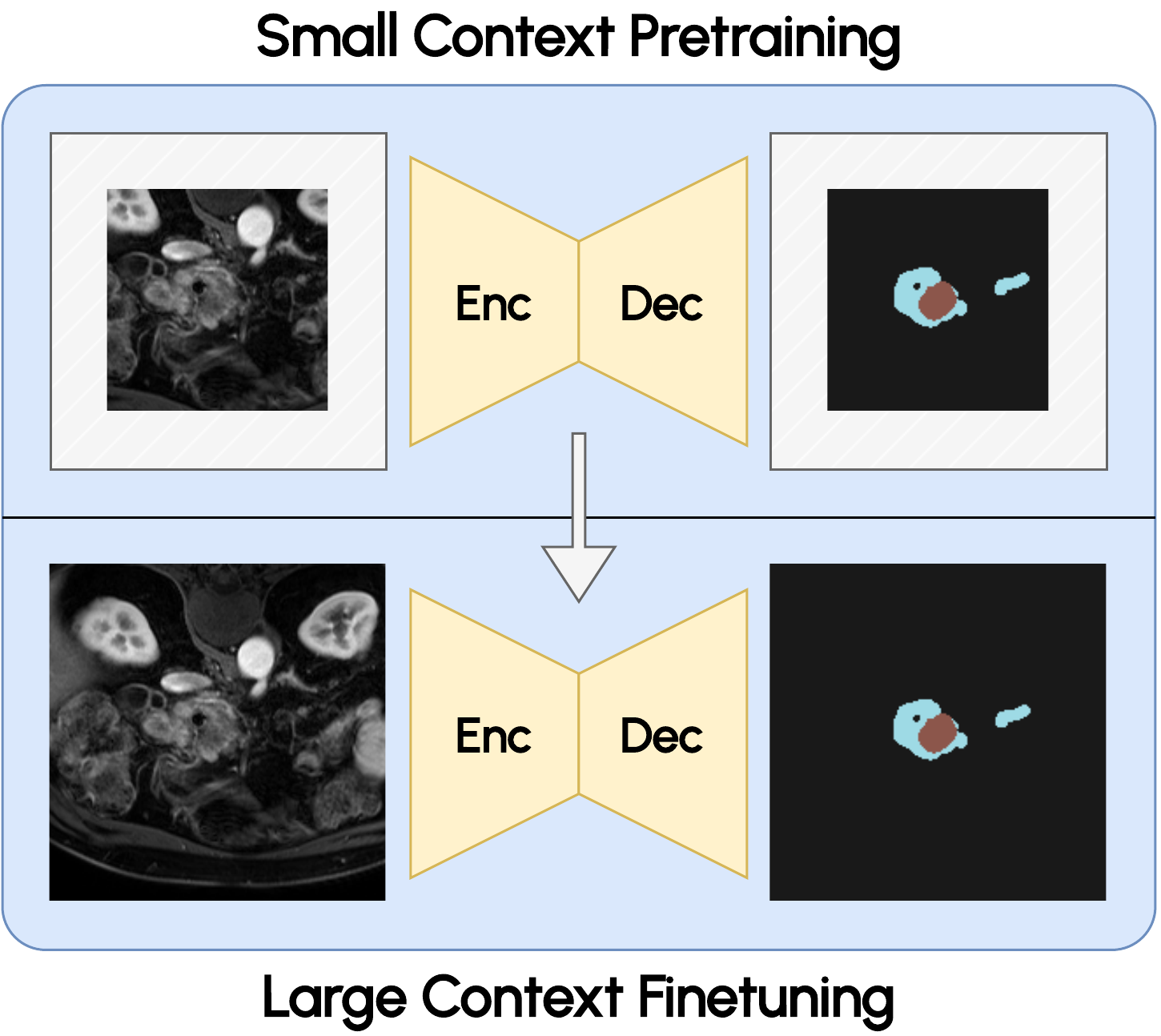}
        \caption{Context scaling strategy}
        \label{fig:sub2}
    \end{subfigure}    
    \caption{\textbf{Network Improvements.} Our network scaling targets the base number of channels (\textcolor{green}{C}) while \textcolor{blue}{\textbf{3D GRN}} improves the micro architecture by limiting activation saturation or collapse during training. Also shown is our context scaling strategy.}
    \label{fig:block_design}
\end{figure}

\subsection{Scaling data only after validating backbones}
\label{sec:scaling-after}
One of our interesting findings is that 9 out 11 large-scale methods that we have explored in \cref{tab:systemic} do not perform any validation of the backbone before introducing data scaling. This introduces performance-based challenges during downstream fine-tuning of a backbone, especially when fine-tuning for clinical purposes leveraging automated segmentation such as radiotherapy planning \cite{putz2025segment} or target localization in biopsies \cite{soerensen2021deep} -- namely, a weak backbone will be restricted to weaker pretrained representations. In this work, we propose validating our backbone against existing state of the art architectures on 4 public validation datasets -- BTCV \cite{landman2015miccai}, AMOS22 \cite{ji2022amos}, KITS21 \cite{heller2023kits21} and ACDC \cite{bernard2018deep} (more in \cref{appx:validation-datasets}). We choose the 5 architectures used in backbones in \cref{tab:systemic} and 3 state-of-the-art architectures from recent medical image segmentation benchmarks \cite{bassi2024touchstone,isensee2024nnu}. 

As demonstrated in \cref{tab:benchmarking_backbones} based on Dice Similarity Coefficient (DSC) that MedNeXt \cite{roy2023mednext} is the best performing network \textit{from literature}, with ResEncL \cite{isensee2024nnu} being a close second. In fact, ResEncL is considered to be a state-of-the-art architecture in literature based on its performance \cite{bassi2024touchstone}, but is only used by 2 out 11 large-scale pretrained nets in \cref{tab:systemic}. This highlights that -- \textbf{a)} Regular benchmarking (or consulting \textit{latest} benchmarking initiatives) for backbones prior to data scaling should be encouraged for better large-scale pretraining, \textbf{b)} MedNeXt is an effective architecture for further development, as shown by the best performance of our newly introduced MedNeXt-v2 in the same benchmark in \cref{tab:benchmarking_backbones}.



\subsection{MedNeXt-v2: Compound scaling ConvNeXts}
MedNeXt (henceforth referred to as MedNeXt-v1) is a popular ConvNeXt-based architecture for 3D medical image segmentation effective in representation learning on small-scale as well as large scale training. In essence, it is designed as a 5-layer UNet \cite{ronneberger2015u} with ConvNeXt blocks at every layer of the network (further in \cref{appx:sec:mednextv2-arch}) . Our goal is to sensibly scale the architecture and training data to obtain a MedNeXt-v2 variant capable of learning stronger representations in 3D medical image datasets. We also aim to enable finetuning of this model at a fractional of the training compute, thereby allowing rapid adoption in downstream segmentation tasks. MedNeXt-v1 leveraged the idea of compound scaling from EfficientNet \cite{tan2019efficientnet} to simultaneously scaling multiple components of the network. We further modify the network micro-architecture and scale along these lines, as described in the following sections.

\subsubsection{3D Global Response Normalization (GRN)}
\label{sec:grn}
2D GRN was a micro-architectural component introduced in the ConvNeXt-V2 \citep{woo2023convnext} in the natural image domain to prevent feature collapse in training ConvNeXt networks. The position of these blocks post the expansion layer encourages the learning of diverse representations by preventing overly dominant feature maps, as the expansion ratio is increased. We adopt it as a 3D GRN block with $l$2 normalization for a volumetric feature map $X_i = \gamma * X_i * \mathcal{N}(X_i) + \beta + X_i \in \mathcal{R}^{H \times W \times D}$
where $\mathcal{N}(X_i) = \frac{||X_i||}{\Sigma_{j=1}^C ||X_j||} \in \mathcal{R}$ for $C$-channels and $\beta, \gamma$ are learnable parameters. We incorporate GRN in all 3D MedNeXt blocks across our architecture for effective pretraining and fine-tuning performance. Additionally, we ablate our choice of 3D GRN on the same settings as our backbone validation in \cref{tab:grn-results}, as well as the effect on activations following fine-tuning in \cref{tab:grn-results}, further demonstrating the effectiveness of the module.

\begin{table}[h]
\centering
\caption{\textbf{3D GRN improves segmentation performance in MedNeXt-v2.} 3D GRN in our MedNeXt-v2 stabilizes performance on average (\textit{always} best or second-best) and gives an edge over both MedNeXt-v1 and SOTA baselines when training from-scratch, leading to higher gains following large-scale pretraining (\cref{fig:v2_better_than_v1}). \textbf{Bold}: Best results, \underline{Underline}: Second-best results.}
\label{tab:grn-results}
\begin{adjustbox}{width=0.475\textwidth}
\begin{tabular}{rccccc}
\textbf{Model} & \textbf{BTCV} & \textbf{AMOS} & \textbf{KiTS} & \textbf{ACDC} & \textbf{Mean} \\
\hline
nnUNet & 79.84 & 87.95 & 87.67 & 93.34 & 87.20 \\
ResEncL & 80.66 & \textbf{88.12} & 88.25 & \textbf{93.44} & 87.62 \\
\midrule
MedNeXt-v1 & \textbf{81.02} & 88.02 & \underline{88.84} & 93.08 & 87.74 \\
\rowcolor{green!10} MedNeXt-v2 & \underline{80.99} & \underline{88.05} & \textbf{89.09} & \underline{93.39} & \textbf{88.03} \\
\bottomrule
\end{tabular}
\end{adjustbox}
\end{table}

\subsubsection{Scaling of ConvNeXt Nets}
We work to establish the idea of scaled networks paired with large-pretraining data. We investigate varying degrees of scaling enabled by the MedNeXt-v2 micro architecture. They are as follows:
\begin{itemize}
    \item \textbf{Depth Scaling:} We adopt the 52 layer depth of the original large (L) variant of MedNeXt-v1 for the default depth level of our v2 architecture. We maintain this scaling \textit{throughout all our experiments}.
    \item \textbf{Width Scaling:} Standard ResNet-like blocks allow for increases in channel size.
    We perform width scaling via increasing the base channel size to $2C$, thereby doubling capacity, while maintaining the $R$ to values as in MedNeXt-v1 \cite{roy2023mednext}.
    \item \textbf{Context Scaling:} 3D deep segmentation networks typically leverage patch-based training for limiting VRAM consumption (similar to other tasks in medical image analysis \citep{ciresan2012deep,hou2016patch,roy20222,roy2025investigating}), where patch size limits the degree of available context for the deep network. While smaller patch sizes of $96\times 96 \times 96$ or $128 \times 128 \times 128$ \cite{lee20223d, hatamizadeh2021swin, roy2023mednext} have become commonplace, this limits performance on certain tasks. We scale the patch size to $192 \times 192 \times 192$ as one of our scaling investigations to increase available context to our network during fine-tuning.
\end{itemize}

\paragraph{Limiting Input Context during pretraining.} There is a dichotomy in scaling of input context by increasing patch size in 3D medical images. On the one hand, it is fundamentally beneficial to leverage larger input context for better performance. However, using large patch sizes in large-scale 3D architectures is memory-intensive and requires significant compute. However, we find it effective to use the standard input patch size of $128 \times 128 \times 128$ in pretraining, and scale it up to $192 \times 192 \times 192$ during fine-tuning. This is effective as our pipeline favors rapid fine-tuning (300 epochs) which is 20\% of the pretraining schedule.

\subsection{Data-Scaling via Large Scale Pretraining}
\label{sec:pretraining}
As mentioned in \cref{sec:pretraining}, large-scale pretraining has been one of the drivers of computer vision research over the last decade \cite{sun2017revisiting, ridnik2021imagenet}. Accordingly, to enable generalization across diverse imaging domains with varying texture, contrast and noise characteristics, we pretrain our model on 18k images of the publicly available subset of the CADS data collection \cite{xu2025cads}. We select a subset of 44 target structures from the available labels thereby simplifying the granularity of the original dataset while ensuring wide anatomical coverage to learn effective generalized representations without optimizing for fine-grained details. This is suitable given our intention to fine-tune our network on downstream tasks.


\paragraph{Pretraining strategy} 
For pretraining, we use the nnUNet framework \cite{isensee2021nnu} to optimize our model across 4 A100 GPUs with distributed data parallel training. The models are trained for 1500 epochs with 250 batches per epoch, a batch size of 8 (2 per GPU) and a patch size of $128 \times 128 \times 128$. We use the AdamW optimizer \cite{loshchilov2017fixing}  used in the original MedNeXt \citep{roy2023mednext} with linear weight decay and an initial learning rate of 1e-3. We use the default nnUNet preprocessing, data augmentations, loss etc in the course of the pretraining. 

\paragraph{Fine-tuning strategy}
The relatively recent introduction of large-scale architectures in 3D medical image segmentation often sees them used in conjunction with optimizers which do not specifically account for convergence stability early in training -- for example, SGD in \citep{isensee2024nnu,huang2023stu} with networks as large as 1.4B Parameters and 12.6 TFLOPS. To stabilize our fine-tuning, we add Linear Warmup \cite{liu2019variance} of 50 epochs to the AdamW optimizer with an maximum learning rate of 1e-3 and fine-tune for a total of 300 epochs with 250 random batches of size 2 per epoch. We use a patch size of $128 \times 128 \times 128$ unless otherwise stated.

\section{Experimental Setup}
\subsection{Datasets}
\paragraph{Initial Benchmarking Datasets}
We use 4 datasets as explored in \citep{isensee2024nnu} for our backbone benchmarking and GRN validations prior to data scaling. Our datasets consist of diverse target structures, imaging modalities and training set sizes - 1) Beyond-The-Cranial-Vault (BTCV) organ segmentation CT dataset \citep{landman2015miccai}, 2) AMOS Organ Segmentation (AMOS22) CT Dataset \citep{ji2022amos}, 3) MICCAI Kidney Tumor Segmentation Challenge 2023 (KiTS23) \citep{heller2023kits21} CT dataset and 4) Automatic Cardiac Diagnostic Challenge (ACDC) \citep{bernard2018deep} CINE MR dataset to develop our methods, with 30, 200, 489 and 200 samples respectively. We use z-score normalization on all datasets and resample them to isotropic spacing. Results on these datasets are expressed as Dice Similarity Coefficient (DSC) as in \cref{tab:benchmarking_backbones,tab:grn-results} on a 80-20 data split strategy for training and validation sets.

\paragraph{Final Evaluation Datasets}
We use 5-fold cross validation for our final evaluation providing Dice Similarity Coefficient (DSC) and Normalized Surface Distance (NSD) at 1mm tolerance on a diverse and challenging set of six public datasets across CT and MR modalities to perform our evaluation. Similar to our initial backbone benchmarking, we use z-score normalization on all datasets and resample them to isotropic spacing. They are described as follows:
\begin{enumerate}
    \item \textit{Pediatric CT-Seg~\cite{pediatric-ct-seg} (D1):} Pediatric CT exams with expert organ contours from multiple institutions and scanners.
    \item \textit{Stanford Knee MR~\cite{stanford_knee} (D2):} Manual segmentations of six knee tissues: patellar and femoral cartilage, lateral/medial tibial cartilage, and lateral/medial meniscus.
    \item \textit{Toothfairy~\cite{ToothFairy3_2024TMI,ToothFairy3_2024IEEEACCESS} (D3):} CBCT volumes with 77-class dental segmentations, including teeth, canals, jaws, and implants.
    \item \textit{Stanford Brain Mets~\cite{stanfordBrainMets} (D4):} brain MRI studies with radiologist-annotated brain metastases segmentations across multiple sequences.
    \item \textit{PANTHER Pacreatic Tumor~\cite{panther} (D5):} T1-weighted contrast-enhanced pancreatic MRIs with pancreas and tumor annotations.
    \item \textit{CTSpine1k~\cite{deng2024ctspine1klargescaledatasetspinal} (D6):} Large-scale spine CT dataset  expert-labeled vertebrae. We exclude all samples from the popular Liver Tumor Segmentation \cite{antonelli2022medical} task from this dataset to prevent overlaps with pretraining cohorts of AbdomenAtlas1.0 in \cref{tab:main_results}.
\end{enumerate}

\subsection{Large-scale Pretrained Baselines}
We use seven publicly available pretrained methods for both benchmarking their performance against each other and as a baseline against our pretrained MedNeXt-v2 architectures. We use TotalSegmentator (TotalSeg) \cite{wasserthal2023totalsegmentator}, MRSegmentator (MRSeg.) \cite{akinci2025totalsegmentator}, CADS \cite{xu2025cads}, the segmentation backbone of Vista3D \cite{he2025vista3d}, the pretrained ViT backbone of SegVol \cite{du2024segvol} incorporated in an UNETR decoder \cite{hatamizadeh2022unetr}, STU-Net \cite{huang2023stu} and the original MedNeXt-v1 \cite{roy2023mednext}. Only the last 2 among these 7, namely STU-Net and MedNeXt-v1, were not introduced accompanied by a large-scale pretraining dataset and were pretrained on the AbdomenAtlas1.0 dataset \cite{qu2023abdomenatlas} as part of the Touchstone large-scale medical image segmentation benchmark \cite{bassi2024touchstone}. The remaining 5 were trained on the same datasets they accompanied when they were proposed. We use the fine-tuning strategy as described in \cref{sec:pretraining} for all methods for fair comparison against MedNeXt-v2 with a patch size of $128^3$ similar to our base MedNeXt-v2. However, we maintain the nnUNet-proposed large patch size of CADS on each dataset for fairer comparison as its ResEncL backbone depends on increased patch size for performance.


\subsection{Scaled MedNeXt-v2 variants}
We experiment with three variants of our MedNeXt-v2 representing different forms of scaling which are as follows:

\begin{enumerate}
    \item \textbf{Base Network:} This is the base variant of our proposed MedNeXt-v2 architecture which follows the large (L) configuration of the MedNeXt-v1 with an added GRN module for improved representation learning.
    \item \textbf{Patch $\times$ 1.5:} The base network performs pretraining and fine-tuning with a patch size of $128 \times 128 \times 128$. In this variant of the network, we scale the patch side length by 1.5 times to $192 \times 192 \times 192$ for the fine-tuning schedule.
    \item \textbf{Width $\times$ 2.0:} In this variant, the number of channels of the base network is increased by a factor of two at each stage of the network. This effectively doubles network capacity of the wide configuration compared to the base network.
\end{enumerate}

\begin{figure}[h]
    \centering
    \begin{minipage}[t]{0.32\linewidth}
        \centering
        \small \textbf{Input + GT}
        \scalebox{1}[1]{\includegraphics[width=\linewidth, trim=0cm 0cm 0cm 0cm, clip]{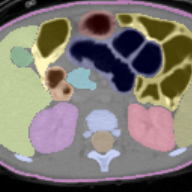}}
    \end{minipage}
    \begin{minipage}[t]{0.32\linewidth}
        \centering
        \small \textbf{v2 (GRN)}
        \scalebox{1}[1]{\includegraphics[width=\linewidth, trim=0cm 0cm 0cm 0cm, clip]{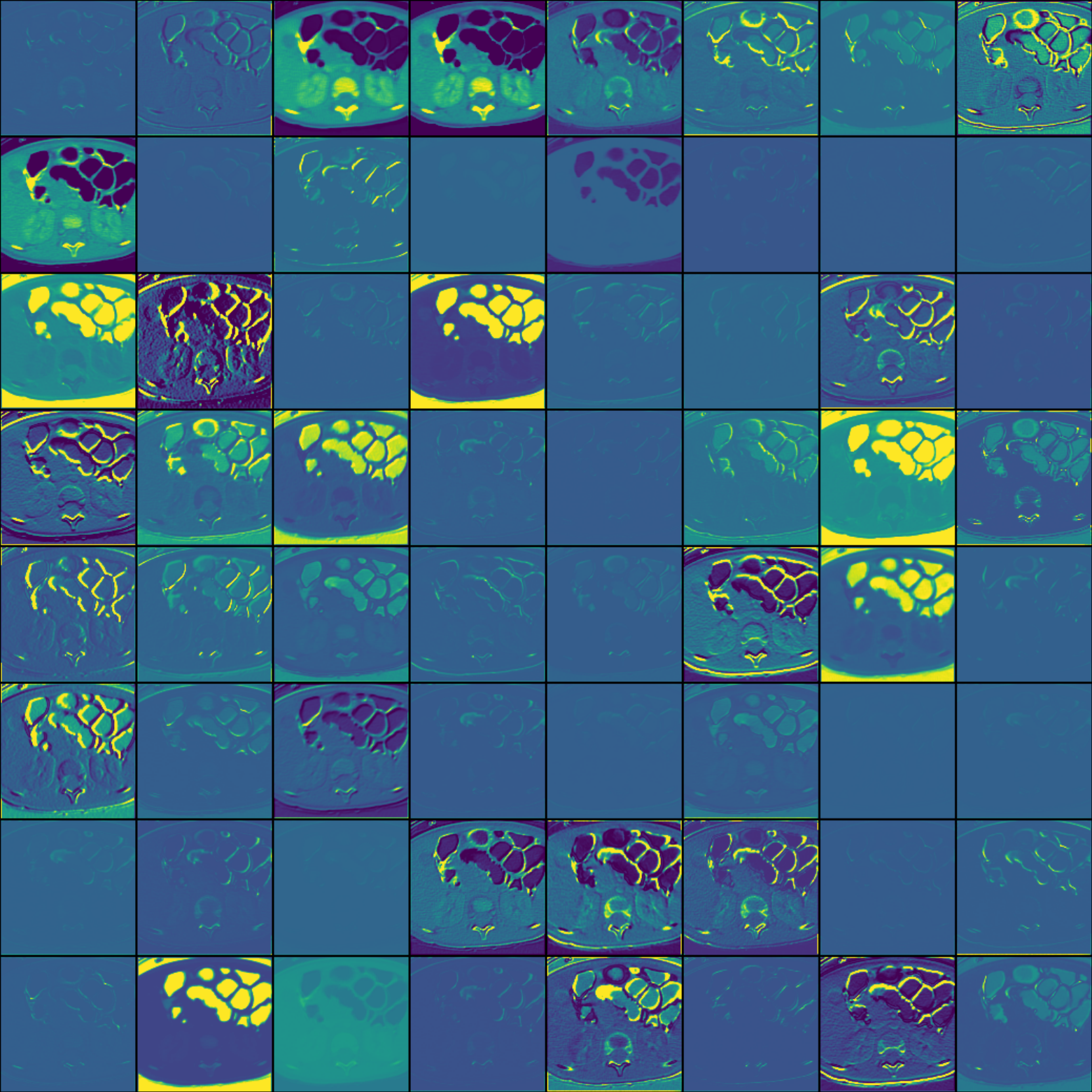}}
    \end{minipage}
    \begin{minipage}[t]{0.32\linewidth}
        \centering
        \small \textbf{v1 (no GRN)}
        \scalebox{1}[1]{\includegraphics[width=\linewidth, trim=0cm 0cm 0cm 0cm, clip]{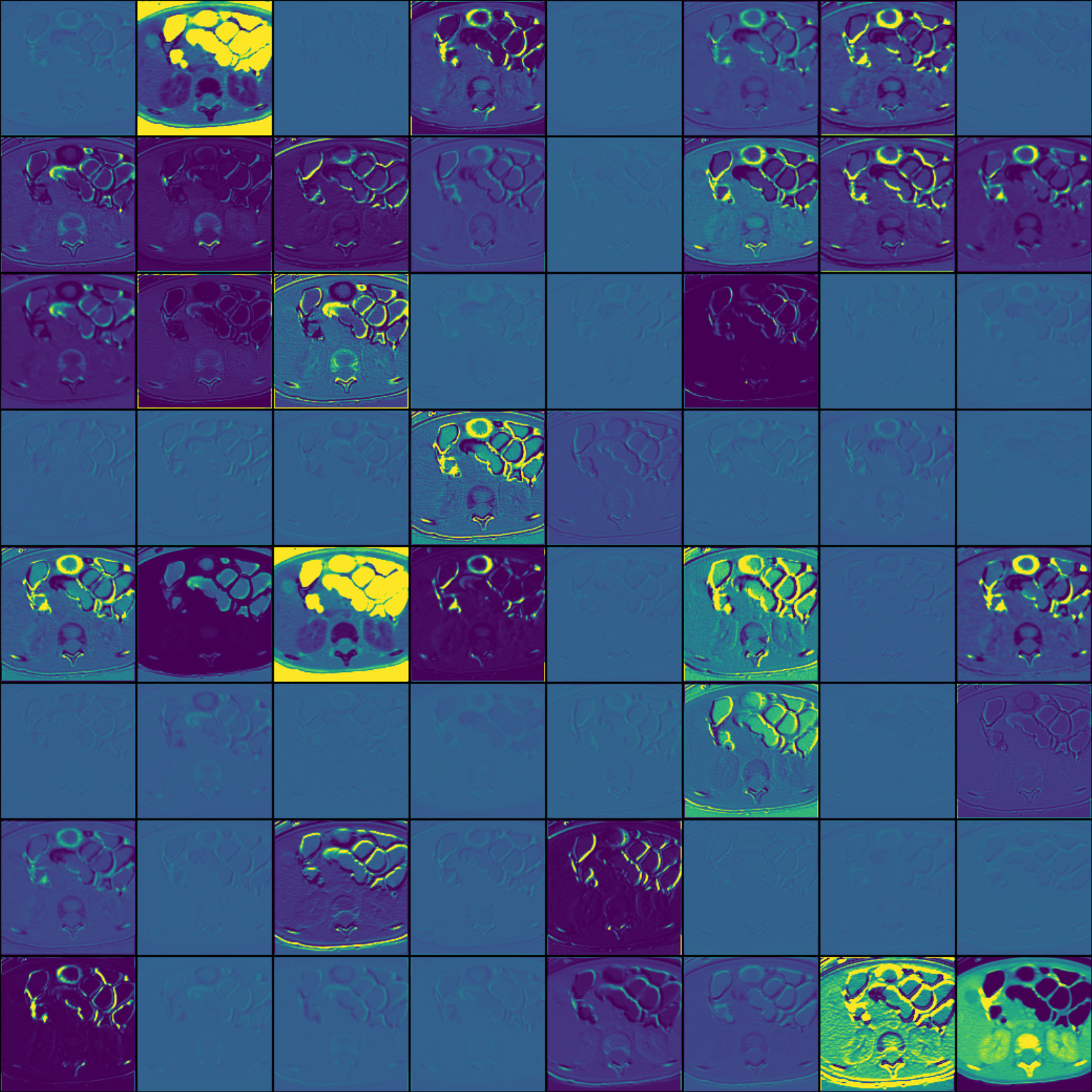}}
    \end{minipage}

    \begin{minipage}[t]{0.32\linewidth}
        \centering
        \scalebox{1}[1]{\includegraphics[width=\linewidth, trim=0cm 0cm 0cm 0cm, clip]{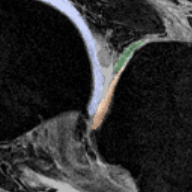}}
    \end{minipage}
    \begin{minipage}[t]{0.32\linewidth}
        \centering
        \scalebox{1}[1]{\includegraphics[width=\linewidth, trim=0cm 0cm 0cm 0cm, clip]{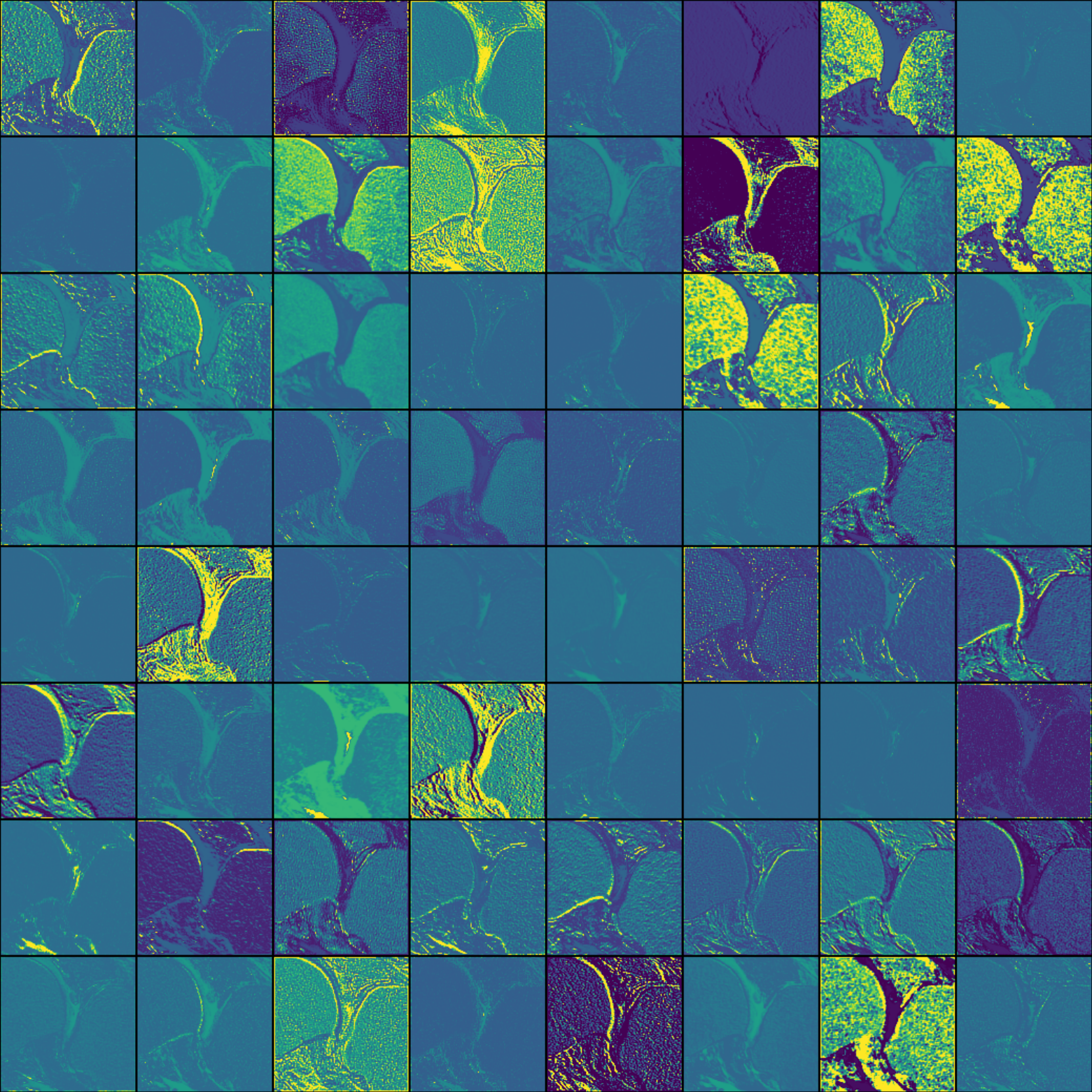}}
    \end{minipage}
    \begin{minipage}[t]{0.32\linewidth}
        \centering
        \scalebox{1}[1]{\includegraphics[width=\linewidth, trim=0cm 0cm 0cm 0cm, clip]{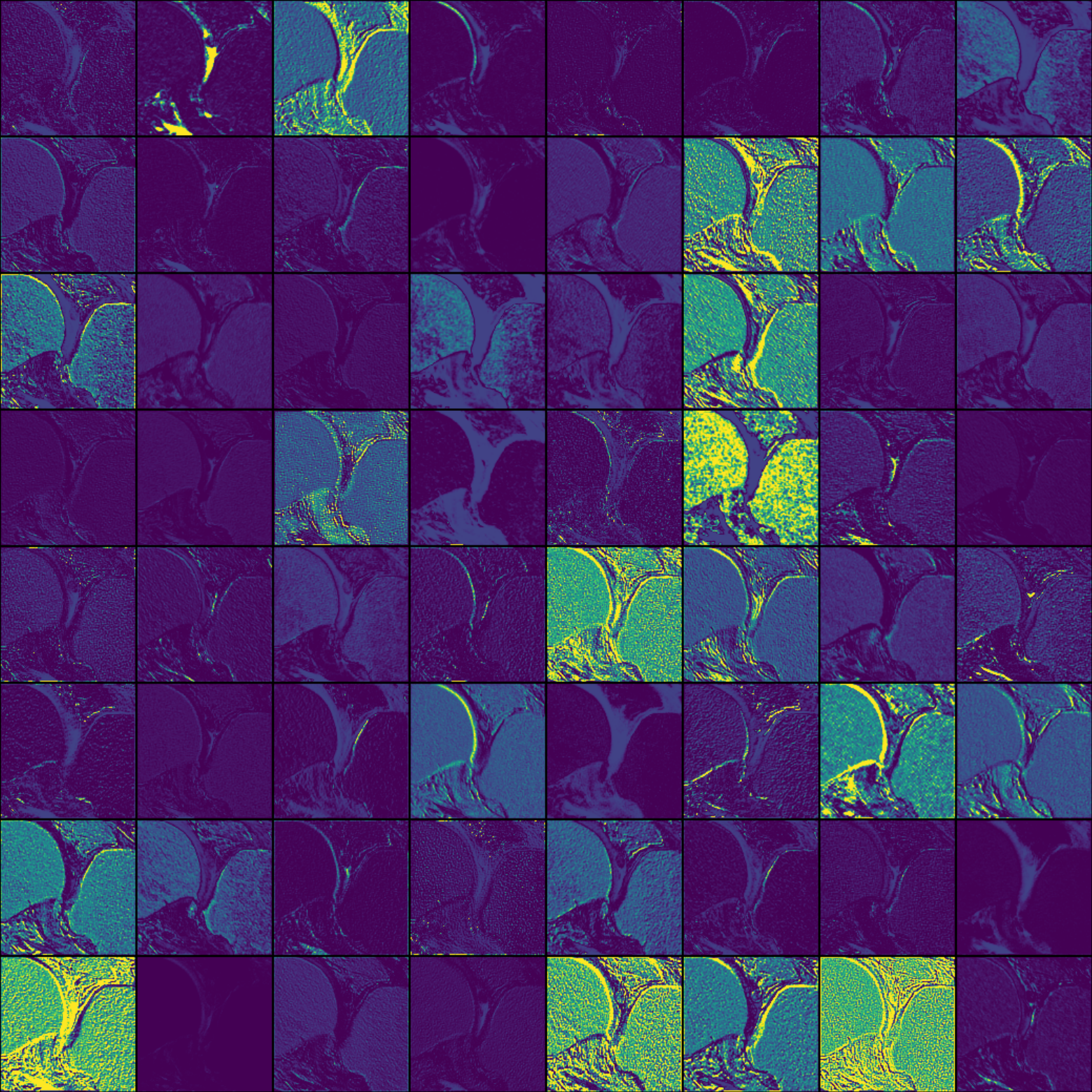}}
    \end{minipage}
    
    \begin{minipage}[t]{0.32\linewidth}
        \centering
        \scalebox{1}[1]{\includegraphics[width=\linewidth, trim=0cm 0cm 0cm 0cm, clip]{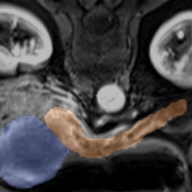}}
    \end{minipage}
    \begin{minipage}[t]{0.32\linewidth}
        \centering
        \scalebox{1}[1]{\includegraphics[width=\linewidth, trim=0cm 0cm 0cm 0cm, clip]{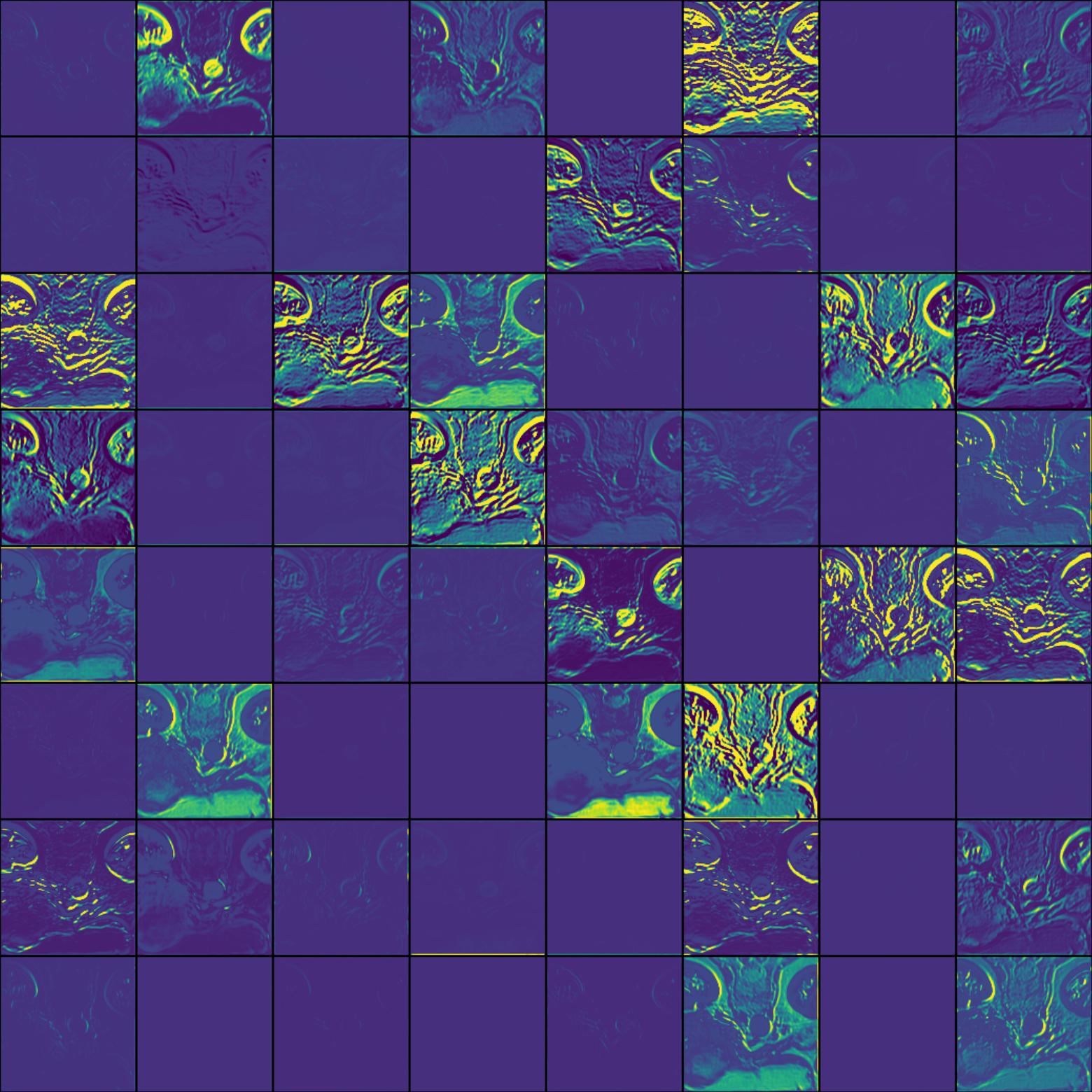}}
    \end{minipage}
    \begin{minipage}[t]{0.32\linewidth}
        \centering
        \scalebox{1}[1]{\includegraphics[width=\linewidth, trim=0cm 0cm 0cm 0cm, clip]{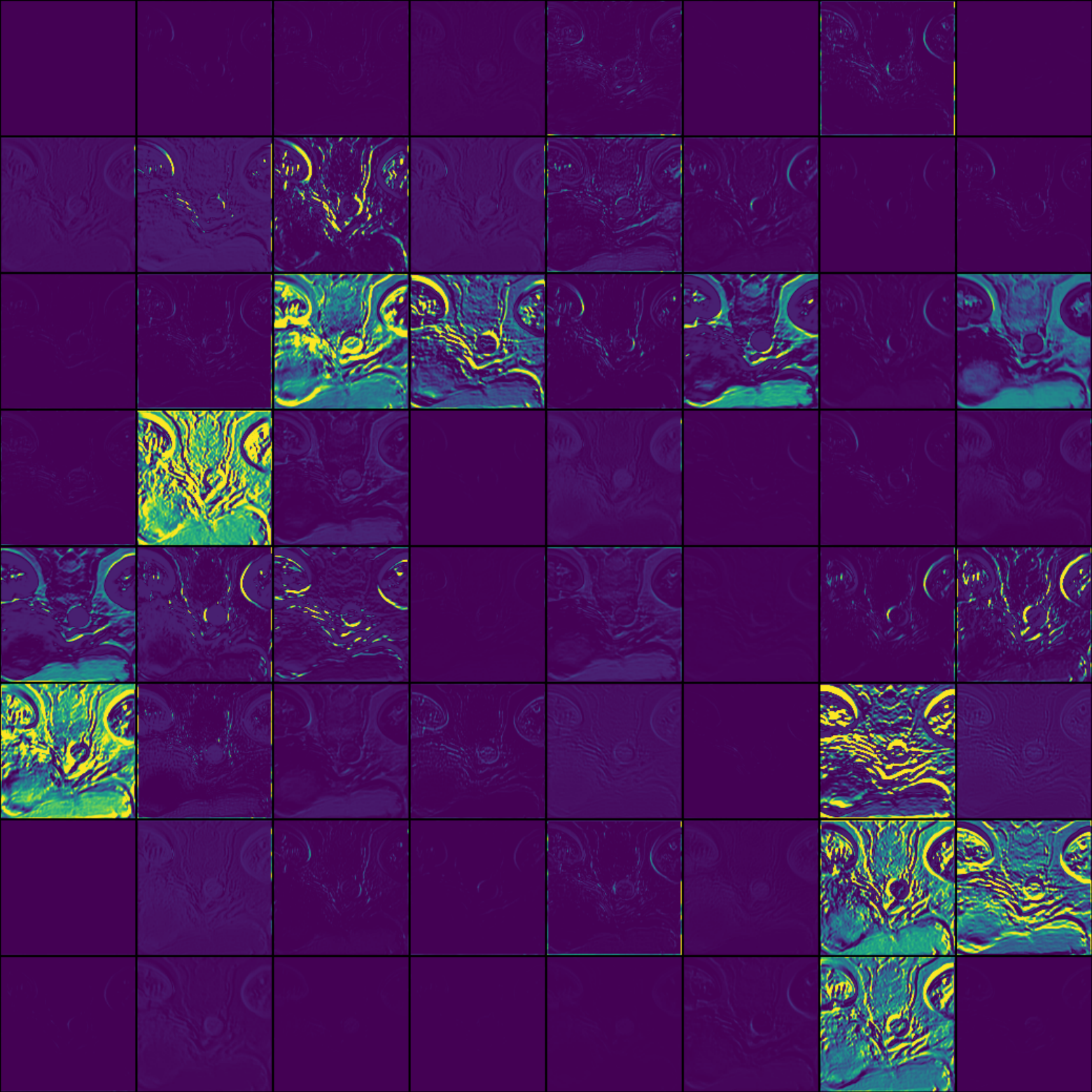}}
    \end{minipage}


    \caption{\textbf{Channel activation visualization demonstrates that 3D GRN reduces redundant activations.} Akin to ConvNeXt-v2 \cite{woo2023convnext}, our visualization of 64 activations in layer 1 of MedNeXt-v2 (with GRN) and MedNeXt-v1 (without GRN) on  Pediatric-CT (D1), Stanford Knee (D2) and Pancreatic Tumor (D5) from \cref{tab:main_results} demonstrates that 3D GRN prevents dead or saturated activations in 3D medical image segmentation tasks, preventing feature collapse and aiding representation learning.}
    \label{fig:grn-vs-nogrn}
\end{figure}


\begin{table*}
\centering
\caption{\textbf{MedNeXt-v2 demonstrates state-of-the-art performance across 144 structures against existing pretrained baselines.} We validate our methods on 6 diverse datasets (3 CT and 3 MR) across 144 anatomical and pathological structures. We demonstrate state-of-the-art segmentation performance based on both DSC and NSD against 7 large-scale pretrained networks. We are also the first to systematically benchmark large-scale pretrained networks at scale and offer significant insights for such models (\cref{sec:benchmarking_analysis}). \textbf{Bold}: Best results, \underline{Underline}: Second-best results.
}
\label{tab:main_results}
\begin{adjustbox}{width=0.975\textwidth}
\rowcolors{4}{gray!15}{white}
\begin{tabular}{r|c|ccccccc} 
\toprule
\rowcolor{gray!15} \textbf{Pretrained} & \textbf{Pretraining} & \multicolumn{7}{c}{\textbf{Dice Similarity Coefficient (DSC)}} \\
\rowcolor{gray!15} \textbf{Models} & \textbf{Datasets} & \textbf{D1} & \textbf{D2} & \textbf{D3} & \textbf{D4} & \textbf{D5} & \textbf{D6} & \textbf{Mean} \\ 
\midrule
\multicolumn{9}{c}{\textbf{Training from Scratch}}  \\
\hline
\textbf{nnUNet} \cite{isensee2021nnu} & - & 79.81/74.93 & 87.28/87.37 & 85.63/88.89 & 65.37/77.23 & 68.70/45.31 & 96.63/97.22 & 80.57/78.49 \\
\textbf{ResEncL} \cite{isensee2024nnu} & - & 82.62/78.39 & 87.03/87.06 & 88.55/91.47 & 65.67/77.81 & 69.11/46.05 & 96.89/97.57 & 81.65/79.72 \\
\textbf{MedNeXt-v1} \cite{roy2023mednext} & - & 84.28/80.22 & 87.26/87.47 & 89.38/92.29 & 64.87/77.73 & 69.37/46.49 & \underline{97.12}/\underline{97.74} & 82.05/80.32 \\
\textbf{MedNeXt-v2} & - & \underline{84.84}/\underline{80.74} & 87.16/87.33 & 89.49/92.40 & 64.89/77.04 & 70.41/46.81 & 97.06/97.70 & 82.31/80.34 \\
\midrule
\multicolumn{9}{c}{\textbf{Pretrained Baselines (using Publicly Available Weights)}} \\
\midrule
\textbf{TotalSeg.} \cite{wasserthal2023totalsegmentator} & TotalSeg-CT & 77.60/72.19 & 87.19/87.17 & 86.74/90.18 & 67.85/80.21 & 71.74/46.83 & 96.11/96.60 & 81.20/78.86 \\
\textbf{MRSeg.} \cite{akinci2025totalsegmentator} & TotalSeg-MR & 78.41/73.19 & 87.29/87.32 & 87.54/90.98 & 67.02/79.17 & 71.62/46.50 & 96.21/96.72 & 81.35/78.98 \\
\textbf{CADS} \cite{xu2025cads} & CADS-Organ & 81.74/77.27 & 87.08/87.16 & \underline{90.34}/\underline{93.37} & 68.07/79.69 & 71.32/47.04 & 96.75/97.36 & 82.55/80.31 \\
\textbf{Vista3D} \cite{he2025vista3d} & Vista & 78.78/73.70 & 87.18/87.13 & 87.75/91.30 & 65.85/77.89 & 70.03/45.83 & 96.44/96.95 & 81.01/78.80 \\
\textbf{SegVol} \cite{du2024segvol} & SegVol & 73.95/65.98 & 85.63/84.37 & 74.19/78.15 & 51.10/62.08 & 60.24/34.71 & 89.10/88.56 & 72.37/68.97 \\
\textbf{STUNet-L} \cite{huang2023stu} & Abd.Atlas1.0 & 80.70/76.18 & 87.12/87.18 & 88.61/91.82 & 66.15/77.69 & \underline{71.99}/47.53 & 96.40/96.94 & 81.83/79.56 \\
\textbf{MedNeXt-v1} \cite{bassi2024touchstone} & Abd.Atlas1.0 & 82.02/77.81 & \underline{87.33}/\underline{87.55} & 89.09/92.03 & \underline{68.52}/80.21 & 71.01/47.88 & 96.81/97.39 & 82.47/80.52 \\
\midrule
\multicolumn{9}{c}{\textbf{MedNeXt-v2 (Our Pretrained Weights)}} \\
\midrule
\textbf{Base Network} & CADS-sub & 83.15/79.12 & 87.28/87.45 & 89.29/92.43 & \textbf{69.04}/\textbf{81.23} & 71.93/\underline{48.49} & 97.03/97.62 & \underline{82.95}/\underline{81.06} \\
\textbf{Patch $\times$ 1.5} & CADS-sub & \textbf{85.59}/\textbf{81.72} & 87.15/87.26 & \textbf{91.77}/\textbf{94.59} & 68.12/\underline{80.26} & \textbf{72.38}/\textbf{48.93} & \textbf{97.20}/\textbf{97.87} & \textbf{83.70}/\textbf{81.77} \\
\textbf{Width $\times$ 2.0} & CADS-sub & 83.62/79.81 & \textbf{87.35}/\textbf{87.57} & 89.44/92.44 & 67.03/79.42 & 71.94/48.12 & 97.04/97.68 & 82.74/80.84 \\
\bottomrule
\end{tabular}
\end{adjustbox}
\end{table*}

\section{Results and Discussion}

\subsection{MedNeXt-v2 is state-of-the-art for 3D medical image segmentation}
\label{sec:mednext-v2}
MedNeXt-v2 leverages micro-architectural improvements and data scaling to demonstrate state-of-the-art performance for 3D medical image segmentation. We discuss the nuances of our performance against MedNeXt-v1 as well as seven other large-scale pretrained baselines. We also discuss the merits of using higher network capacity and larger input context during finetuning in the following sections.

\begin{figure}
    \centering
    \includegraphics[width=0.975\linewidth]{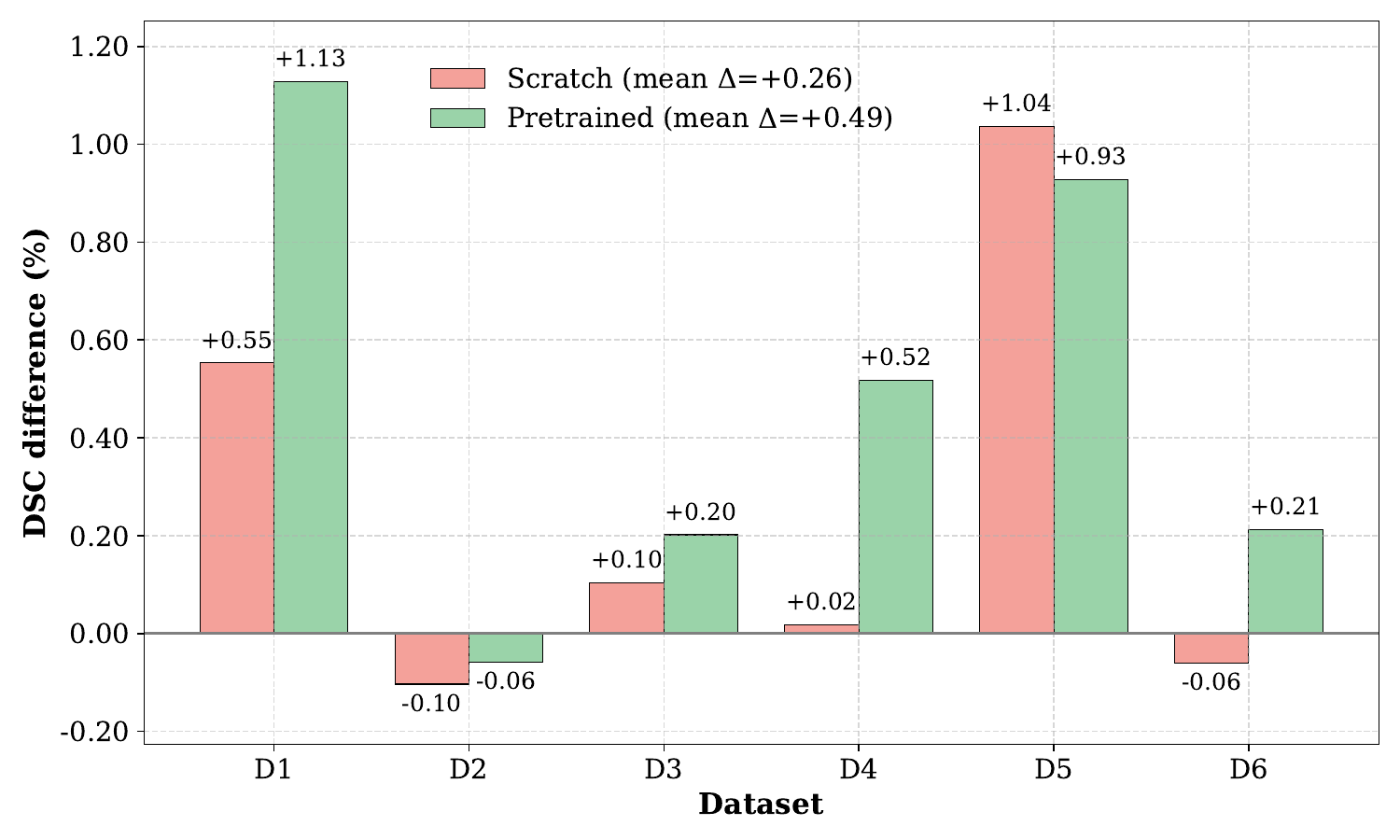}
    \caption{\textbf{MedNeXt-v2 outperforms MedNeXt-v1 from scratch and during finetuning.} The addition of the GRN stabilizes the performance of the v2 architecture and improves performances compared to MedNeXt-v1 as seen in \cref{tab:main_results}. We see improvements $>$1.0 Dice points on tasks as diverse as the segmentation of Pediatric Organs in CTs \textbf{(D1)} and Pancreatic Tumor in MR \textbf{(D5)}. We only observe limited gains on the highly saturated knee segmentation task \textbf{D2} for all methods.}
    \label{fig:v2_better_than_v1}
\end{figure}

\paragraph{Improvement over MedNeXt-v1.}  
As illustrated in \cref{fig:v2_better_than_v1}, the integration of a GRN module delivers consistent improvements over MedNeXt-v1, with and without pretraining. This is owing to a similar phenomenon as in ConvNeXt-v2 of preventing feature collapse represented by dead or saturated neuron which we visualize in \cref{fig:grn-vs-nogrn}. Across \textit{144 structures} on 6 challenging public datasets spanning CT and MR modalities, MedNeXt-v2 demonstrates state-of-the-art performance. On challenging datasets for Pediatric CT Segmentation (D1) and pancreatic tumor segmentation (D5), it achieves gains of up to \textbf{+1.0 DSC} despite saturated accuracy ceilings, affirming MedNeXt-v2 as an architecture for effective representation learning in 3D medical image segmentation.

\paragraph{SOTA among pretrained baselines} 
All MedNeXt-v2 variants are seen to outperform \textbf{\textit{all seven large-scale pretrained baselines}} in \cref{tab:main_results}. This includes average gains of \textbf{+2.69 DSC, +2.97 NSD} and \textbf{+1.15 DSC, +1.46 NSD} respectively over powerful pretrained networks like Vista3D and CADS, as illustrated in \cref{fig:qualitative-main}, demonstrating MedNeXt-v2 as a state-of-the-art large-scale pretrained backbone. In particular, our outperforming of CADS on similar pretraining data demonstrates the benefits of the ConvNeXt architecture as a superior backbone for leveraging large annotated pretraining datasets (see also \cref{para:better_backbones_better_pretraining}) for downstream segmentation tasks across CT as well as MR modalities.

\begin{figure}[ht]
    \centering
    \begin{minipage}[t]{0.32\linewidth}
        \centering
        \small \textbf{Input + GT}
        \scalebox{1}[1]{\includegraphics[width=\linewidth, trim=0cm 0cm 0cm 0cm, clip]{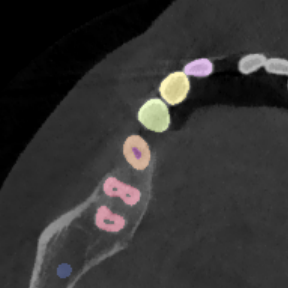}}
    \end{minipage}
    \begin{minipage}[t]{0.32\linewidth}
        \centering
        \small \textbf{v2 (Context: $192^3$)}
        \scalebox{1}[1]{\includegraphics[width=\linewidth, trim=0cm 0cm 0cm 0cm, clip]{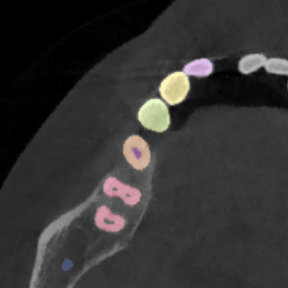}}
    \end{minipage}
    \begin{minipage}[t]{0.32\linewidth}
        \centering
        \small \textbf{v2 (Context: $128^3$)}
        \scalebox{1}[1]{\includegraphics[width=\linewidth, trim=0cm 0cm 0cm 0cm, clip]{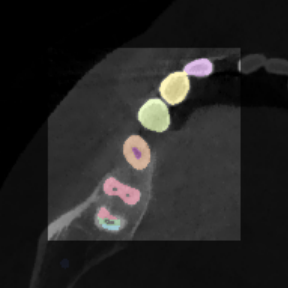}}
    \end{minipage}

    \caption{\textbf{Increased context during fine-tuning improves performance.} Increasing the available spatial context to 3.375 times with $192^3$ patches is a cheap and effective strategy to leverage our pretrained MedNeXt-v2 while limiting pretraining costs. Importantly, we see an example from Toothfairy (D3) where added spatial context of the jaw enables better segmentation of the teeth near the image boundary, which a MedNeXt-v2 fine-tuned on $128^3$ patches is unable to segment accurately.}
    \label{fig:large-context}
\end{figure}

\begin{figure*}[ht]
    \centering
    \begin{minipage}[t]{0.12\linewidth}
        \centering
        \scalebox{1}[1]{\includegraphics[width=\linewidth, trim=2.5cm 0cm 2.9cm 0cm, clip]{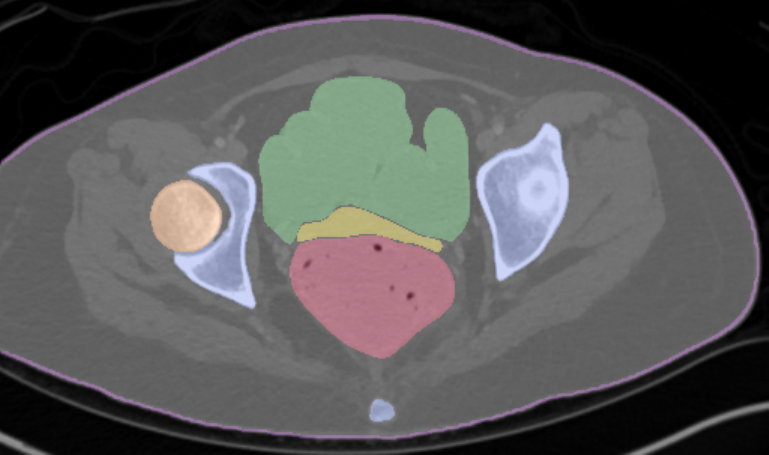}}
    \end{minipage}
    \begin{minipage}[t]{0.12\linewidth}
        \centering
        \scalebox{1}[1]{\includegraphics[width=\linewidth, trim=2.5cm 0cm 2.9cm 0cm, clip]{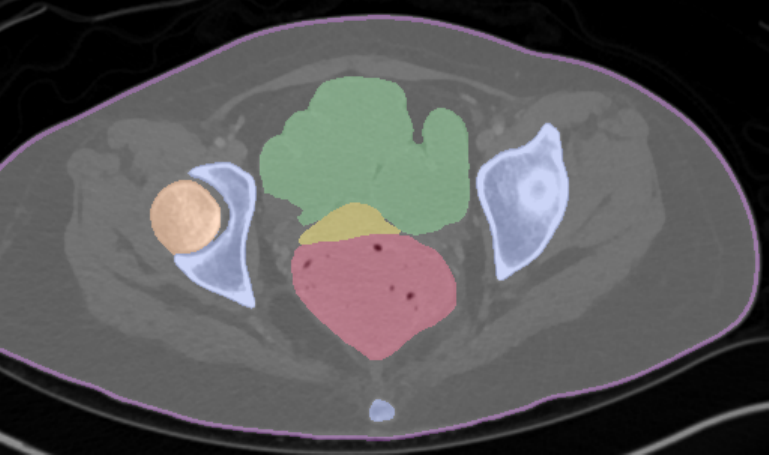}}
    \end{minipage}
    \begin{minipage}[t]{0.12\linewidth}
        \centering
        \scalebox{1}[1]{\includegraphics[width=\linewidth, trim=2.5cm 0cm 2.9cm 0cm, clip]{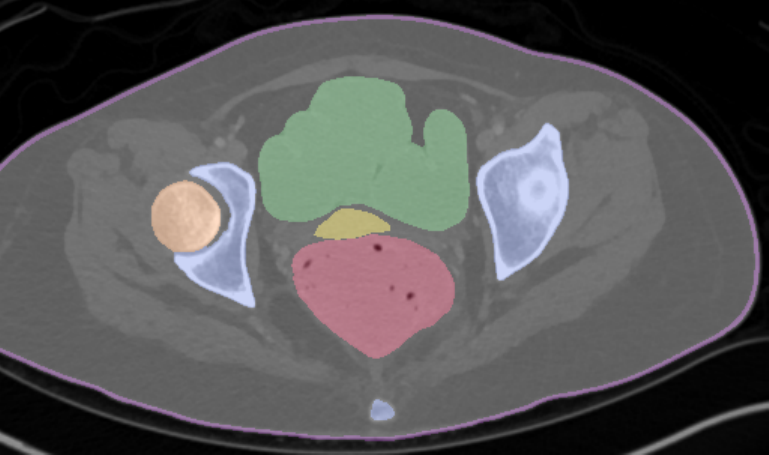}}
    \end{minipage}
        \begin{minipage}[t]{0.12\linewidth}
        \centering
        \scalebox{1}[1]{\includegraphics[width=\linewidth, trim=2.5cm 0cm 2.9cm 0cm, clip]{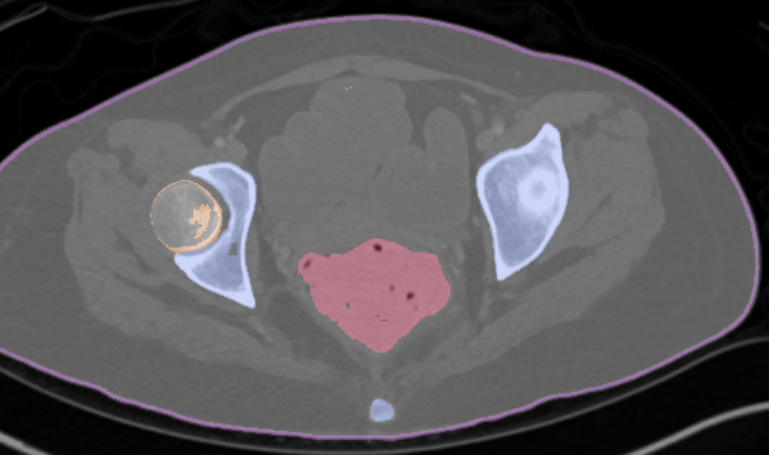}}
    \end{minipage}
    \begin{minipage}[t]{0.12\linewidth}
        \centering
        \scalebox{1}[1]{\includegraphics[width=\linewidth, trim=2.5cm 0cm 2.9cm 0cm, clip]{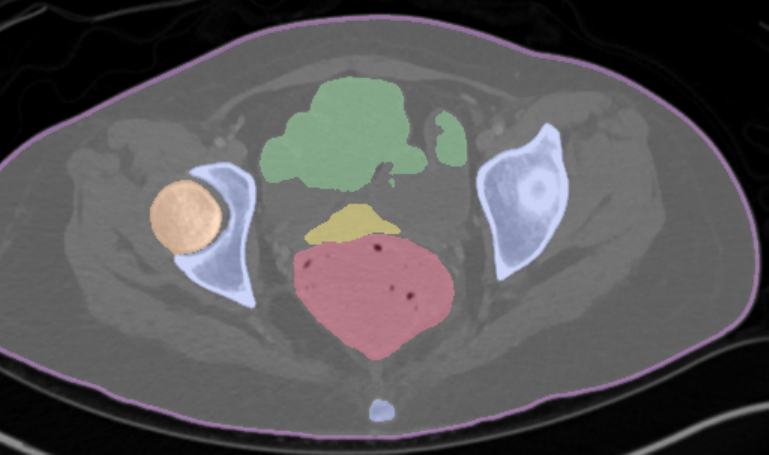}}
    \end{minipage}
    \begin{minipage}[t]{0.12\linewidth}
        \centering
        \scalebox{1}[1]{\includegraphics[width=\linewidth, trim=2.5cm 0cm 2.9cm 0cm, clip]{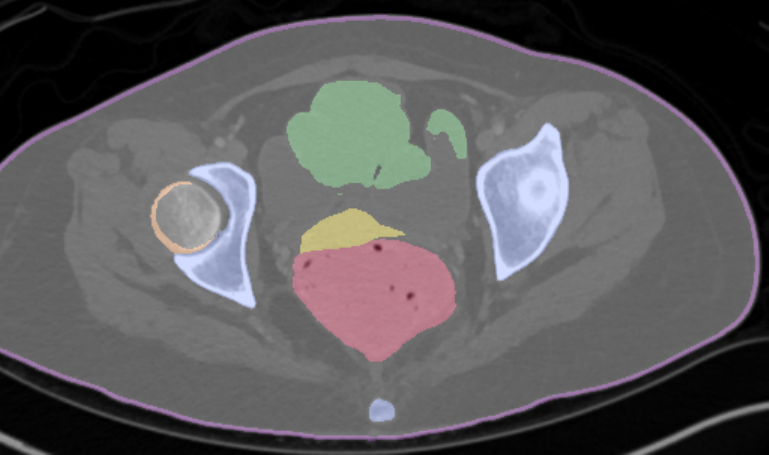}}
    \end{minipage}
    \begin{minipage}[t]{0.12\linewidth}
        \centering
        \scalebox{1}[1]{\includegraphics[width=\linewidth, trim=2.5cm 0cm 2.9cm 0cm, clip]{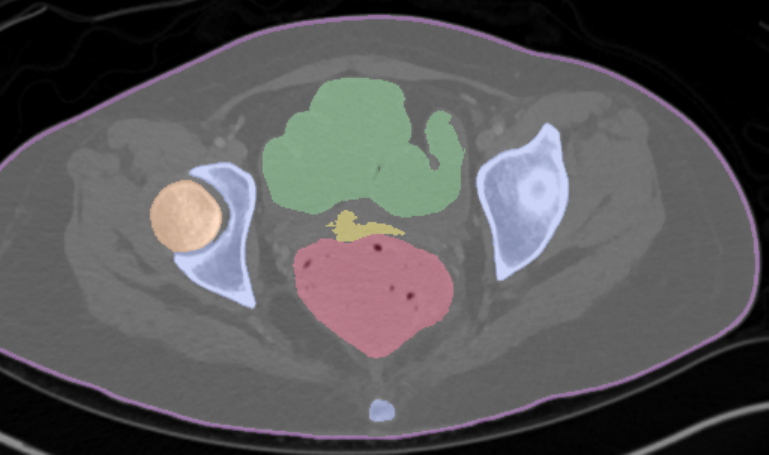}}
    \end{minipage}
    \begin{minipage}[t]{0.12\linewidth}
        \centering
        \scalebox{1}[1]{\includegraphics[width=\linewidth, trim=2.5cm 0cm 2.9cm 0cm, clip]{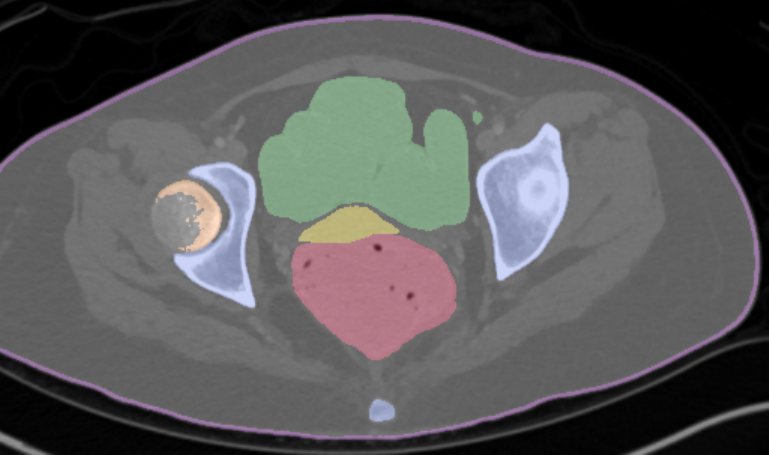}}
    \end{minipage}

    \begin{minipage}[t]{0.12\linewidth}
        \centering
        \scalebox{1}[1]{\includegraphics[width=\linewidth, trim=0cm 0.5cm 0cm 0cm, clip]{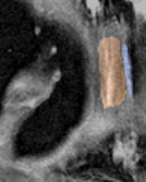}}
        \small \textbf{Input + GT}
    \end{minipage}
    \begin{minipage}[t]{0.12\linewidth}
        \centering
        \scalebox{1}[1]{\includegraphics[width=\linewidth, trim=0cm 0.5cm 0cm 0cm, clip]{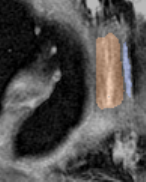}}
        \small \textbf{MedNeXt-v2}
    \end{minipage}
    \begin{minipage}[t]{0.12\linewidth}
        \centering
        \scalebox{1}[1]{\includegraphics[width=\linewidth, trim=0cm 0.5cm 0cm 0cm, clip]{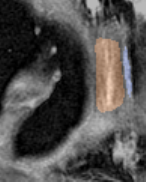}}
        \small \textbf{v2-192}
    \end{minipage}
        \begin{minipage}[t]{0.12\linewidth}
        \centering
        \scalebox{1}[1]{\includegraphics[width=\linewidth, trim=0cm 0.5cm 0cm 0cm, clip]{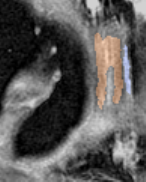}}
        \small \textbf{SegVol}
    \end{minipage}
    \begin{minipage}[t]{0.12\linewidth}
        \centering
        \scalebox{1}[1]{\includegraphics[width=\linewidth, trim=0cm 0.5cm 0cm 0cm, clip]{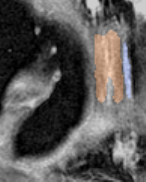}}
        \small \textbf{Vista3D}
    \end{minipage}
    \begin{minipage}[t]{0.12\linewidth}
        \centering
        \scalebox{1}[1]{\includegraphics[width=\linewidth, trim=0cm 0.5cm 0cm 0cm, clip]{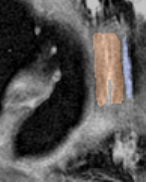}}
        \small \textbf{TotalSeg}
    \end{minipage}
    \begin{minipage}[t]{0.12\linewidth}
        \centering
        \scalebox{1}[1]{\includegraphics[width=\linewidth, trim=0cm 0.5cm 0cm 0cm, clip]{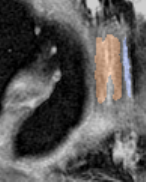}}
        \small \textbf{CADS}
    \end{minipage}
    \begin{minipage}[t]{0.12\linewidth}
        \centering
        \scalebox{1}[1]{\includegraphics[width=\linewidth, trim=0cm 0.5cm 0cm 0cm, clip]{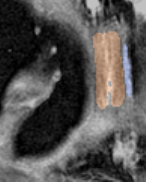}}
        \small \textbf{MedNeXt-v1}
    \end{minipage}
    
    \caption{\textbf{Qualitative visualizations.} We visualize 1 CT (D1: Pediatric CT-Seg) and 1 MR (D2: Stanford Knee MR) dataset to demonstrate the effectiveness of MedNeXt-v2 and MedNeXt-v2 with large $192^3$ context against other pretrained baselines including MedNeXt-v1.}
    \label{fig:qualitative-main}
\end{figure*}

\paragraph{Scaling context during finetuning} 
Our increase in side length of the input patch to $192^3$ increases the spatial context available to MedNeXt-v2 by 3.375$\times$. This results in a successful scaling strategy and performance is seen to improve in numerous datasets leveraging this spatial context. In particular, we offer an illustration in \cref{fig:large-context} on Toothfairy dataset (D3) where this extra context is seen to help with the segmentation of a structure (\textit{tooth}) by possibly leveraging the spatial position of the nearby anatomy (\textit{jaw}). Quantitatively, we see large improvements on D3 of \textbf{+1.43 DSC} and \textbf{+3.57} on D1 (Pediatric CT-Seg) against other pretrained methods. While our large input patch variant MedNeXt-v2 performs the best in our benchmark, we do not attribute any unfair advantage to our network. CADS, for example, leverages a large input patch size as well but does not show performance as strong as MedNeXt-v2. The combination of a strong backbone and larger input patches benefits MedNeXt-v2 in achieving strong downstream performance.

\paragraph{Scaling Capacity}
\label{sec:scaling-mednextv2-capacity}
We obtain mixed results with scaling capacity (Width $\times 2.0$) during pretraining and fine-tuning. Doubling the base channel capacity of the 52-layer base network increases parameters from 62M to 247M and its usage during pretraining massively scales required compute. However, while we see superior performance on average to all seven existing pretrained baselines during finetuning on both DSC and NSD, it does not exceed the performance of the smaller MedNeXt-v2 base network or the MedNeXt-v2 with larger input patch scaling. Coupled with the limited performance of 440M parameter STU-Net, this motivates us to reflect on naively scaling network capacity in the face of existing pretraining dataset sizes in \cref{sec:arch_scale_diminish}.

\subsection{Benchmarking Analysis}
\label{sec:benchmarking_analysis}
Our benchmarking of existing large-scale backbones for 3D medical image segmentation offers significant insights and constitutes one of the first systematic evaluations of fine-tuning these architectures.

\paragraph{Better backbones lead to better pretraining}
\label{para:better_backbones_better_pretraining}
In our backbone validation in \cref{tab:benchmarking_backbones}, we hypothesized that better performance from scratch would lead to improvements in pretraining (and consequently during finetuning). We demonstrate that this is indeed the case when comparing the performance of CADS (ResEncL) and MedNeXt-v2 in \cref{tab:main_results}.  Both are similarly pretrained, but the performance from-scratch is carried through with MedNeXt-v2 outperforming the CADS (ResEncL) model, by margins as large as \textbf{+1.0 DSC} on segmenting small ($<$ 0.05cm$^3$) metastatic lesions \cite{wald2025primus} in brain MRIs (D4). This is not restricted to a single modality as similar performance (\textbf{+1.4 DSC}) is also seen when segmenting organs in pediatric CTs (D1), with equivalent gains in NSD indicating accurate surface delineation. This highlights that the choice of backbone is key to taking advantage of large pretraining datasets for accurate downstream segmentation performance.

\begin{figure}[h]
    \centering
    \includegraphics[width=0.95\linewidth]{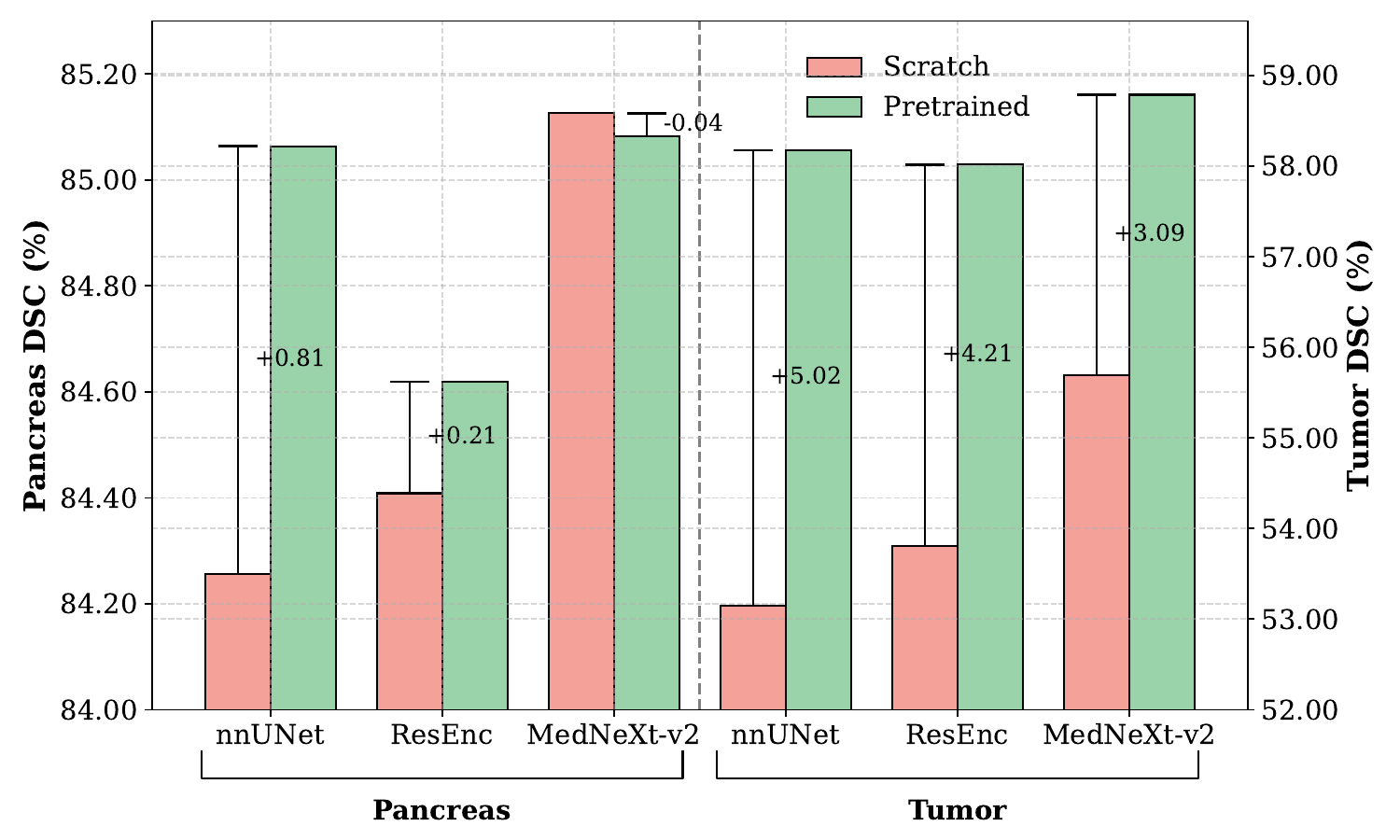}
    \caption{\textbf{Tumor classes benefit massively in pretraining.} Our analysis on Pancreatic Tumor Segmentation dataset (D4) in \cref{tab:main_results} demonstrate that across multiple networks, the benefit of pretraining in accurate tumor segmentation is significantly more than for the segmentation of the pancreas.}
    \label{fig:tumor-vs-notumor}
\end{figure}

\paragraph{General Pretraining vs Modality-Specific Pretraining}
A key finding from our benchmark is the comparison between TotalSeg-CT and MR, which use CT and MR pretraining, respectively, while sharing a similar nnU-Net backbone. Pretraining yields substantial gains, exceeding \textbf{+2.0 DSC} on brain metastasis (D4) and pancreatic tumor segmentation (D5) compared to training from scratch. However, the two models remain nearly indistinguishable across CT and MR evaluation datasets in terms of DSC and NSD. This raises the question of the effectiveness of \textit{isolated} modality-specific pretraining, suggesting that for structurally similar modalities (e.g., CT, MR, PET-CT), learned representations generalize well across modalities, thereby making the effect of modality-specific pretraining inseparable from standard pretraining. 

\paragraph{Tumors vs Healthy Structures}




To better understand the role of pretraining in the segmentation of pathological versus non-pathological structures, we use pancreatic tumor segmentation (D5) as a representative case. We compare three pretrained models, namely TotalSegmentator-CT (nnU-Net), CADS (ResEncL), and MedNeXt-v2 (base) against their from-scratch counterparts, as shown in \cref{fig:tumor-vs-notumor}. In doing so, we observe two consistent trends:

\begin{itemize}
\item Across all backbones, pretraining yields substantial improvements on the \textit{tumor class}, with MedNeXt-v2 achieving the highest absolute performance but the smaller nnU-Net backbone exhibiting the largest relative gain of approximately \textbf{+5.0 DSC}.
\item For \textit{organ segmentation}, results are more mixed with the larger networks showing moderate to no gains following pretraining. However, smaller backbones are again seen to benefit the most.
\end{itemize}

These findings suggest that \textit{pathological segmentation tasks benefit disproportionately from large-scale pretraining}, and may in fact serve as more discriminative benchmarks for evaluating generalization than the segmentation of healthy anatomical structures.



\section{Conclusion}
Developing deep learning architectures that perform reliably across the wide variety of 3D medical image segmentation tasks \cite{isensee2021nnu} remains challenging. Although training from scratch has dominated the field over the past decade, methodological advances often focused on gains in small-scale data regimes. In this work, we embrace the shift toward large-scale supervised representation learning and introduce MedNeXt-v2, a pretrained ConvNeXt-based architecture that achieves state-of-the-art performance in 3D medical image segmentation. We validate our design against seven pretrained baselines and derive several key insights from what is, to our knowledge, one of the first systematic evaluations of publicly available pretrained methods with an emphasis on downstream finetuning. We hope that our findings, along with our open-source code and pretrained models, will foster further research in this direction.

\section*{Acknowledgements}
The present contribution is supported by the Helmholtz Association under the joint research school "HIDSS4Health – Helmholtz Information and Data Science School for Health". This work was partly funded by Helmholtz Imaging (HI), a platform of the Helmholtz Incubator on Information and Data Science.

The authors also gratefully acknowledge the computing time provided on the high-performance computer HoreKa by the National High-Performance Computing Center at KIT (NHR@KIT). This center is jointly supported by the Federal Ministry of Education and Research and the Ministry of Science, Research and the Arts of Baden-Württemberg, as part of the National High-Performance Computing (NHR) joint funding program (https://www.nhr-verein.de/en/our-partners). HoreKa is partly funded by the German Research Foundation (DFG). This work was supported by the Helmholtz Association's Initiative and Networking Fund on the HAICORE@FZJ partition.

Finally, we would like to wholeheartedly acknowledge the help of our colleagues at the German Cancer Research Center (DKFZ),  namely, Balint Kovacs, Katharina Eckstein and Moritz Langenberg for their invaluable help towards the completion of this manuscript.

\newpage
{
    \small
    \bibliographystyle{ieeenat_fullname}
    \bibliography{main}
}

\newpage
\appendix
\maketitlesupplementary

\section{Large-scale pretrained models in literature}
There have been a plethora of deep learning architectures with large-scale supervised pretraining in 3D medical image segmentation. We use seven such architectures in our benchmarking in \cref{tab:main_results}. The models are described briefly with their backbone architecture as follows:

\begin{enumerate}
    \item \textbf{TotalSegmentator:} TotalSegmentator is an nnUNet \cite{isensee2021nnu} architecture trained on 1228 CT images with 117 anatomical structures. 
    \item \textbf{MRSegmentator:} MRSegmentator is an nnUNet architecture trained on 616 volumes with 50 anatomical structures. The project is jointly maintained as part of TotalSegmentator.
    \item \textbf{CADS:} CADS presents a ResEncL nnUNet \cite{isensee2024nnu} trained on a collection of 22022 CT volumes. While each volume contains 167 annotated structures, the pretrained weights are made available for major anatomical subsets of structures. We use the pretrained weights for subset 1 defined as ``\textit{major abdominal organs, primary thoracic organs (lungs), and major abdominal vasculature}" \cite{xu2025cads} in the course of this work.
    \item \textbf{Vista3D:} Vista3D uses a modified SegResNet backbone trained on 11454 CT volumes with 127 classes, designed to perform both automatic and interactive segmentation. For the purposes of this work, we retrained the segmentation branch while deactivating the interactive branch.
    \item \textbf{SegVol:} SegVol uses a 3D ViT image encoder which is pretrained on 96000 CT volumes followed by finetuning on 6000 CTs on 47 anatomical areas. The ViT is utilized as part of an UNETR image encoder following instruction on the SegVol Github repository. 
    \item \textbf{STUNet-L:} STU-Net uses a modified and significantly scaled up nnUNet variant as the backbone of its work. The pretrained weights are derived from the weights made publicly available by the authors following their participation in the Touchstone benchmark \cite{bassi2024touchstone}.
    \item \textbf{MedNeXt-v1:} MedNeXt-v1 used in the pretrained model section of \cref{tab:main_results} has been pretrained on AbdomenAtlas1.0 \cite{qu2023abdomenatlas} which consists of 5195 Abdomen CT volumes with 10 annotated abdominal structures.
\end{enumerate}

\section{MedNeXt-v2 architecture}
\label{appx:sec:mednextv2-arch}
\begin{figure}
    \centering
    \includegraphics[width=\linewidth, trim=0cm 0cm 0.6cm 0cm, clip]{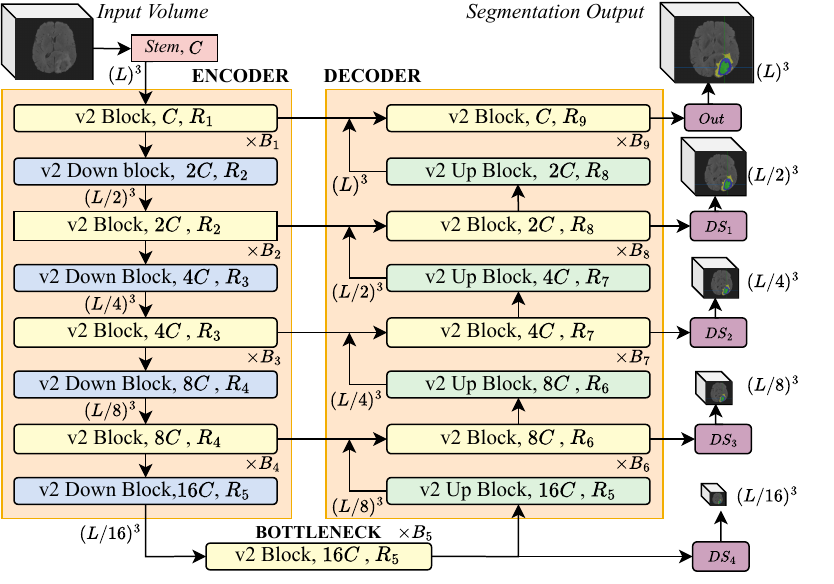}
    \caption{\textbf{MedNeXt-v2 architecture.} The macro-architecure of MedNeXt-v2 is fundamentally similar to that of MedNeXt-v1 \cite{roy2023mednext}. The differences lie in the unique benefits of mirco-architectural changes via the GRN, architecture scaling and leveraging large-scale supervised pretraining for state-of-the-art 3D medical image segmentation.}
    \label{fig:mednextv2-arch}
\end{figure}
The macro architecture of MedNeXt-v2 is similar to that of MedNeXt-v1 \cite{roy2023mednext}. The architecture is symmetrical in terms of encoder and decoder capacities, consists of 5 spatial hierachies, each with $B_i$ blocks and leverages deep supervision at all levels of the decoder. All residuals are additive and all layers benefit from GRN as described in \cref{sec:grn}, including the up and downsampling layers. The layers in \cref{fig:mednextv2-arch} are described as follows:

\begin{enumerate}
    \item \textbf{\textit{Stem}:} The \textit{Stem} is designed to use a Conv3d of kernel size and stride of 1 to project the input channels into the base number of channels ($C$) of the architecture.
    \item \textbf{\textit{MedNeXt-v2 block}:} The \textit{MedNeXt-v2 block} is a residual ConvNeXt block which has the following operations in sequence -- a depthwise convolution layer with convolution kernel size 3, an instance norm layer, a pointwise expansion convolution with an expansion ratio $R$, a GELU activation, a GRN block and a pointwise compression convolution reducing the channels to the base number of channels for the layer. The nomenclature (v2 Block, $4C$, $R_7$) in \cref{fig:mednextv2-arch} indicates that the convolution operation of the block has an input and output of channel count of $4C$ with an expansion ratio of $R_7$ per the architecture configurations (derived from MedNeXt-v1).
    \item \textbf{MedNeXt-v2 Up/Down Block:} The \textit{MedNeXt-v2 Up/Down blocks} (see \cref{fig:block_design}) differ from the standard \textit{MedNeXt-v2 blocks} merely in strided convolutions in the depthwise block with Transposed Convolutions replacing the normal Convolution operation in the upsampling layers. There is also a doubling of channels following downsampling and halving following upsampling as in most UNet architectures. There is also a convolution or strided convolution in the residual connection to adjust the spatial and channel dimensions.
    \item \textbf{\textit{Deep Supervision (DS) / Output (Out) Block:}} Each DS or Out block is merely a convolution of kernel size 1 mapping the number of channels of the input tensor to the number of output classes of the task. 
\end{enumerate}

\section{Backbone Benchmarking models}
We use nnUNet \cite{isensee2021nnu}, SegResNet \cite{myronenko20183d}, ResEncL \cite{isensee2024nnu}, SwinUNETR \cite{hatamizadeh2021swin} and UNETR \cite{hatamizadeh2022unetr} which is the representative popular backbone architecture using a Vision Transformer (ViT) \cite{dosovitskiy2020image}. Competing state-of-the-art methods include MedNeXt (v1) \cite{roy2023mednext}, STU-Net \cite{huang2023stu} and CoTr \cite{xie2021cotr}, where each method is a high performer in recent small and large-scale benchmarks for 3D medical image segmentation \cite{isensee2024nnu, bassi2024touchstone}.

\section{Datasets}
We use a pool of \textit{validation datasets} to validate our backbone prior to architecture scaling (\cref{sec:scaling-after}) on which we report results on a single 80:20 split. We keep these separate from our set of 6 final datasets where we perform more extensive 5-fold cross validation in \cref{tab:main_results}, to avoid optimizing our backbone on our test data.

\subsection{Validation Datasets}
\label{appx:validation-datasets}
Our validation dataset pool consists of four popular public dataset (3 CT and 1 cineMRI) as used in \cite{roy2023mednext,isensee2024nnu}. They are described briefly in the following:

\begin{enumerate}
    \item \textit{BTCV:} The \textit{Beyond-The-Cranial-Vault (BTCV)} is a small but popular dataset consisting of 30 CT images with 13 annotated abdominal organs.
    \item \textit{AMOS22:} The \textit{AMOS22} competition dataset consists of 200 abdominal CT volumes with 15 annotated organs. This is the version of the dataset used prior to the completion of the AMOS22 Grand Challenge and has been used to retain parity with MedNeXt-v1.
    \item \textit{KiTS23:} The \textit{Kidney Tumor Segmentation (KiTS23)} challenge 2023 dataset consists of 489 volumes of 2 annotated labels of kidney and tumor.
    \item \text{ACDC:} The ACDC dataset consists of 200 samples of cineMRI volumes with 3 annotated labels for heart structures. 
\end{enumerate}

\section{The diminishing returns of scaling in 3D medical image segmentation}
\label{sec:arch_scale_diminish}
Following up on our network scaling experiments with MedNeXt-v2 (\cref{sec:scaling-mednextv2-capacity}), we observe that certain datasets, such as Stanford Knee MR (D2), exhibit substantial saturation. The standard deviation across all methods is 0.43 DSC and 0.8 NSD, excluding the lowest-performing SegVol model, these values drop to 0.10 DSC and 0.17 NSD. In fact, there is barely any difference between nnU-Net trained from scratch and any pretrained model, regardless of scale and especially in terms of DSC which focuses on volumetric accuracy. Importantly, this phenomenon occurs in only one of the six datasets in our final evaluation pool. This strongly suggests that the benefits of model scaling and pretraining, while generally positive, cannot reliably overcome inherent performance saturation present in some datasets. As a result, careful consideration is needed when selecting evaluation data as well as depending on just a single metric for large-scale pretrained models, since saturated datasets may lead to inconclusive assessments of pretraining efficacy in future benchmarks.

This is also demonstrated in mixed performance of some datasets with the 30M parameter TotalSegmentator model sometimes having similar finetuning performance as the 440M parameter STU-Net model -- importantly, both share the same ConvNet-based nnUNet backbone. However, the 247M parameter MedNeXt-v2 (Width $\times$ 2.0) model is often noticeably better than both TotalSegmentator and STU-Net, which indicates that the architecture design (ConvNet vs ConvNeXt) might also be a factor in architecture scaling. Merely scaling the network parameters, while sometimes effective, might lead to saturation of performance but can be overcome with better backbone architecture design.

\end{document}